\documentclass[useAMS,usenatbib]{mnras}
\usepackage{amsmath}	
\usepackage{graphicx, textcomp}
\usepackage{ marvosym }
\usepackage{amsmath}
\usepackage{amssymb}	

\usepackage{deluxetable}

\newcommand{\ergscm}{erg cm$^{-2}$ s$^{-1}$}

\newcommand{\dg}{$^\circ$}

\newcommand{\er}{$\pm$}

\title[Modelling the broad-band SED of hotspots]{Particle acceleration in low-power hotspots: modelling the broad-band spectral energy distribution}

\author[G. Migliori et al.]
{G. Migliori$^{1,2}$\thanks{E-mail: giulia.migliori@inaf.it},
M. Orienti$^{1}$,  
L. Coccato$^3$, G. Brunetti$^1$,
F. D'Ammando$^{1}$, 
\newauthor K.-H. Mack$^{1}$, M.A. Prieto$^{4}$\\
$^{1}$Istituto di Radioastronomia - INAF, Via P. Gobetti 101, I-40129 Bologna, Italy\\
$^{2}$Dipartimento di Fisica e Astronomia, Universit\`a di Bologna, Via Gobetti 93/2, I-40129 Bologna, Italy\\
$^{3}$European Southern Observatory, Karl-Schwarzschild-Stra\Shilling e 2, D-85748 Garching b. M\"unchen, Germany\\
$^4$Instituto de Astrof\'{i}sica de Canarias, c/ V\'{i}a L\'actea s/n,
E-38205  La Laguna (Tenerife), Spain\\
}
\date{Received \today; accepted ?}

\pagerange{\pageref{firstpage}--\pageref{lastpage}} \pubyear{2002}

\def\LaTeX{L\kern-.36em\raise.3ex\hbox{a}\kern-.15em
    T\kern-.1667em\lower.7ex\hbox{E}\kern-.125emX}

\begin{document}

\label{firstpage}

\maketitle

\begin{abstract}
The acceleration and radiative processes active in low-power radio hotspots are investigated by means of new deep near-infrared (NIR) and optical VLT observations, complemented with archival, high-sensitivity VLT, radio VLA and X-ray {\it Chandra} data. For the three studied radio  galaxies (3C\,105, 3C\,195 and 3C\,227), we confirm the detection of NIR/optical counterparts of the observed radio hotspots. We resolve multiple components in  3C\,227 West and in 3C\,105 South and characterize the diffuse NIR/optical emission of the latter. We show that the linear size of this component ($\gtrsim$4 kpc) makes 3C\,105 South a compelling case for particles' re-acceleration in the post-shock region. Modeling of the radio-to-X-ray spectral energy distribution (SED) of 3C\,195 South and 3C\,227 W1 gives clues on the origin of the detected X-ray emission. In the context of inverse Compton models, the peculiarly steep synchrotron curve of 3C\,195 South sets constraints on the shape of the radiating particles' spectrum that are testable with better knowledge of the SED shape at low ($\lesssim$GHz) radio frequencies and in X-rays. The X-ray emission of 3C\,227 W1 can be explained with an additional synchrotron component originating in compact ($<$100 pc) regions,  such those revealed by radio observations at 22 GHz, provided that efficient particle acceleration ($\gamma\gtrsim$10$^7$) is ongoing. The emerging picture is that of systems in which different acceleration and radiative processes coexist.
\end{abstract}


\section{Introduction}

Hotspots are compact and bright regions typically located at the edge of the lobes of
powerful radio galaxies. In the standard scenario, the hotspots mark the
region where a relativistic 
jet impacts the 
surrounding medium and particles are accelerated by strong shocks
and may radiate up to X-rays. 
The main mechanism at the origin of
X-ray emission from hotspots is still debated. In many cases, the near infrared (NIR) and optical fluxes rule out a single synchrotron radio-to-X-ray component \citep[see e.g.][for a recent compilation]{zhang18}.  Interestingly, there seems to be a connection between the radio luminosity of the hotspots and their X-ray properties.
In fact, in powerful 
hotspots  (with 1.4 GHz luminosities $\gtrsim$10$^{25}$ W Hz$^{-1}$ sr$^{-1}$), like Cygnus\,A \citep{stawarz07},
the X-ray emission is consistent with synchrotron self-Compton radiation (SSC)
from relativistic electrons
\citep[e.g.][]{harris94,harris00,hardcastle04,KS05,werner12}.
However, in
low-power hotspots  (with 1.4 GHz luminosities $\lesssim$10$^{25}$ W Hz$^{-1}$ sr$^{-1}$), SSC radiation would require a large departure from conditions of energy equipartition between particles and magnetic field to reproduce the observed levels of X-ray
emission \citep{hardcastle04}. An alternative process is 
inverse Compton (IC) scattering off the 
cosmic microwave background (CMB) seed photons \citep{Kat03,Tav05}. For this mechanism to be effective,  the plasma in the hotspot must be still relativistic and the region seen under a small viewing angle, two assumptions in contrast with the constraints derived from the observed symmetrical, large-scale morphology of the radio galaxies.
Moreover, one-zone synchrotron-IC models cannot account for the offsets that are often observed between the centroids of the X-ray and radio-to-optical emission \citep{hardcastle07,perlman10,mo12}. A decelerating jet with multiple, radiatively interacting emitting regions has been put forward as a viable solution by  \citet{gk04}. Alternatively, the X-ray emission may be explained in terms of synchrotron radiation from a highly energetic population  of particles, different from that responsible for the radio-to-optical emission \citep[e.g.][]{hardcastle04,hardcastle07,tingay08,mo12,mingo17,mo17}.\\ 
\indent
 The spectral shape of the synchrotron emission from low and high radio power hotspots are also different,  with the former one having higher break frequencies ($\nu_{break}$, i.e. the synchrotron frequencies of the oldest electrons still within the hotspot volume) than the latter ones. 
 This can be explained if the break frequency is related to the magnetic field strength ($B$) in the hotspot volume, so that the electrons producing the optical emission survive a longer time in low radio power hotspots (with correspondingly lower B) than in high radio power hotspots.
 The observed dependence, $\nu_{break}\propto B^{-3}$, is in agreement with theoretical expectations based on the shock-acceleration model \citep{brunetti03}. On the other hand, the scenario of radiation from particles accelerated by a single strong shock at the jet termination  is challenged by a number of observational issues. In several sources, we observe hotspot complexes with multiple bright features surrounded by diffuse emission  \citep{black92,lehay97}. 
In particular, the discovery of optical diffuse
emission extending on kpc-scale can be hardly reconciled with the short radiative lifetimes of optical-emitting
particles \citep[e.g.][]{aprieto97,prieto02,lv99,cheung05,mack09,erlund10,mo12}. This suggests that efficient and spatially distributed
acceleration mechanisms could be active in the post-shock region.
Recent Atacama Large
Millimeter Array (ALMA) observations of the hotspot 3C\,445 South
unveiled highly polarized regions,
suggesting the presence of shocks, enshrouded by unpolarized diffuse
emission, compatible with instabilities and/or projection
effects in a complex shock surface \citep{mo17}.\\
\indent
In this paper we present results on a new multi-band campaign of
four low-power hotspots from the sample presented in \citet{mack09}: 3C~105 South (z=0.089), 3C~227 East and West (z=0.0863) and 3C~195 South (z=0.109). 
The new Very Large Telescope
(VLT) observations were requested with the goal of defining the hotspots' broad-band spectral energy distribution (SED), constrain the emission
mechanisms at work at high-energies and search for diffuse optical emission,  as a possible signature of particle re-acceleration.  Indeed, the selection  criteria of low power hotspots \citep[radio power, redshift and declination, see][]{mack09} set very constraining limits for optical detection, even with the most sensitive telescopes as the VLT, hence the small number of sources. Nonetheless, dedicated studies of a few sources, representative of the entire population, have the  potential to progress our understanding of the particle acceleration and radiative processes active in these structures.\\
\indent
We analyze the new VLT observations in the NIR/optical bands and radio observations
obtained with the Very large Array (VLA). We also
retrieve archival 
{\it Chandra} data in order to extend the study to
X-rays. Two hotspots (namely 3C\,105 South and 3C\,227
West) showed extended optical emission in
earlier VLT images presented in \citet{mack09}, whereas the other two
(3C\,195 South and 3C\,227 East)
were only tentatively detected and the new VLT observations also aimed at
confirming their NIR-optical emission.\\
This paper is organized as follows: in Section 2 we present the
observations and data analysis; results are reported in Section 3; spectral modeling of the best candidates is described in Section 4 and the results discussed in Section 5 and conclusions are drawn in Section 6.\\
\indent
Throughout this paper, we assume the following cosmology: $H_{0} =
71\; {\rm km/s\, Mpc^{-1}}$, 
$\Omega_{\rm M} = 0.27$ and $\Omega_{\rm \Lambda} = 0.73$,
in a flat Universe. The spectral index, $\alpha$, 
is defined as 
$S {\rm (\nu)} \propto \nu^{- \alpha}$.

\section{Observations}

\subsection{Optical and near infrared observations}

Optical and NIR observations of the
hotspots were taken with the VLT in 
Paranal, Chile, under the observing programs 69.B-0544, 072.B-0360
(P.I. Prieto) and 084.B-0362 (P.I. Orienti).
Observations of period 69 were acquired in 2001-2003 using the Infrared Spectrometer And Array Camera (ISAAC, NIR
imaging). Observations of period 72 were acquired in 2003 November-December, using the FOcal Reducer and low dispersion Spectrograph 1 (FORS1, optical imaging). Observations of period 84 were acquired
between 2009 November and 2010 March, using FORS2 and ISAAC (optical and NIR imaging, respectively). 
The observations performed in 2009 and 2010 are presented here for the first time, whereas those taken between 2001 and 2003 were already presented in \citet{mack09} and \cite{mo12} and are here re-analyzed. 
Table \ref{optical_log}
provides further details on the
observing log.\\
The reduction of individual exposures was carried out
using standard tasks of FORS imaging pipeline, executed under the
ESOreflex environment \citep{Freudling13}. Standard reduction
includes bias observations, sky flats to correct for field
illumination and pixel-to-pixel sensitivity variations. The sky
background was computed in regions free from sources and interpolated
over the field of view.  Individual exposures were then coadded with
the {\tt iraf} task {\tt imcombine} using bright sources as reference
for alignment. Astrometric correction, to compensate systematics in the
telescope pointing, was performed using the Two Micron All Sky Survey
(2MASS) point-like source catalogs \citep{Skr06}.  The number of stars used for the astrometric correction goes from a minimum of 5 (for 3C105\,South  in Ks/J/H) up to 40 (3C195\,South  in R and B), resulting in a positional uncertainty within 0.5\arcsec\ in all cases.\\
 Finally, a precise estimation of the residual sky contamination was
computed on the final stacked image on a region close to the source of
interest to compensate for large-scale variations in the field of view
introduced by the instrument pipeline. The standard deviation of the
counts in the sky-selected region was used to estimate the photometric
error due to the sky background.\\
Despite observations of standard star fields were foreseen, some of
our targets missed the relevant calibrations in several bands (B, R, Ks, H, Js for 3C227\,West; B, R, Ks for 3C227\,East; B, R, Js for 3C105\,South and Ks for 3C195\,South). For
those fields, we calibrated our observations using known sources in
the observed field, exploiting the information of the Position and Proper Motions eXtended (PPMX) catalogue
\citep[][ i.e. the only catalogue that contains non-saturated
sources in our fields of view]{roser08}. The standard deviation of the
difference between the measured and tabulated magnitudes was used to
estimate the error on the photometric zeropoint; this was combined to
the photometric error determined above to compute the total
uncertainties of our measurements (see Table \ref{optical_log}).

\begin{table*}
\caption{VLT observations of the radio hotspots.}  
\begin{center}
\begin{tabular}{lclll}
\hline
Hotspot & Date & Instrument, Band & Central wavelength &Photometric error\\
        & (YYYY-MM-DD) &     &($\mu$m)               & (mag) \\ 
\hline
3C\,105 S & 2001-08-20 &  {\tt ISAAC, Ks}       &2.16       &0.02 \\ 
        & 2002-09-24 &  {\tt ISAAC, H}          &1.65       &0.04 \\ 
        & 2009-11-03 &  {\tt ISAAC, Js}         &1.24       &0.17 \\ 
        & 2003-11-26 &  {\tt FORS1, R\_BESS}    &0.657       &0.86 \\ 
        & 2009-11-25 &  {\tt FORS2, b\_HIGH}    &0.440       &0.30 \\        
3C\,195 S & 2009-11-03 &  {\tt ISAAC, Ks}       &2.16       &0.16 \\ 
        & 2003-01-21 &  {\tt ISAAC, H}          &1.65       &0.04 \\ 
        & 2003-12-18 &  {\tt FORS1, R\_BESS}    &0.657       &0.01 \\ 
        & 2003-11-30 &  {\tt FORS1, B\_BESS}    &0.429       &0.02 \\ 
3C\,227 E & 2010-01-17 &  {\tt ISAAC, Ks}       &2.16       &0.13 \\ 
        & 2003-12-18 &  {\tt FORS1, R\_BESS}    &0.657  &0.28 \\ 
        & 2003-12-18 &  {\tt FORS1, B\_BESS}    &0.429       &0.24 \\ 
3C\,227 W & 2001-04-18 &  {\tt ISAAC, Ks}       &2.16       &0.07 \\ 
        & 2009-12-27 &  {\tt ISAAC, H}          &1.65       &0.14 \\      
        & 2009-12-27 &  {\tt ISAAC, Js}         &1.24       &0.11 \\ 
        & 2010-02-13 &  {\tt FORS2, R\_SPECIAL} &0.655       &0.08 \\ 
        & 2010-02-15 &  {\tt FORS2, b\_HIGH}    &0.440       &0.21 \\ 
\hline
\end{tabular}
\end{center}
\label{optical_log}
\end{table*}

\subsection{X-ray observations}
X-ray observations of our targets were publicly available in the NASA's high-energy archive\footnote{\url{https://heasarc.gsfc.nasa.gov/docs/archive.html.}}. Given the need for high angular resolution
mapping, we considered only {\it Chandra} pointings, which allow us to resolve
X-ray structures on sub-arcsecond scales.   
We re-analyzed for consistency the archival {\it Chandra} observations using
up-to-date calibration files. A log table of the X-ray observations is
reported in Table \ref{Xobs}. 
The observations were taken with the ACIS-S array in very faint (VF) mode.
For 3C\,105 and 3C\,227, the southern hotspot and the western hot spot
complex, respectively, were placed near the aim point, on the S3
chip. Due to the radio source angular extension of 3C\,227, the eastern
hotspot fell on a different chip. The observation of 3C\,195 was centered on the X-ray core, however the whole radio structure ($\lesssim$130\arcsec) fits on the S3 chip.
The X-ray data analysis was performed with the {\it Chandra} Interactive
Analysis of Observation (CIAO) 4.9 software \citep{Fru06} using the
calibration files CALDB version 4.7.7. We ran the
\texttt{chandra$\_$repro} reprocessing script, that performs all the
standard analysis steps.   
We checked and filtered the data for the time intervals of background flares. 
For imaging purposes, the two observations of 3C\,227 were merged together.
 By default, the energy-dependent sub-pixel event-resolution (EDSER) algorithm, which improves the ACIS image quality, was applied to all datasets.
We generated smoothed images in full pixel resolution (0\arcsec.492)
and rebinned to a pixel size of 0\arcsec.296 and 0\arcsec.123. 
The spectral analysis was performed with Sherpa \citep{Free01}. We
employed the cstat statistic \citep{cash79} together with the neldermead
optimization method \citep{neldermead65}. The spectra were not rebinned and, if not differently
specified, the background was modeled. Uncertainties are given at the
90 per cent confidence level.  
When the statistics were too low, we used PIMMS to convert the 0.5--7.0 keV net count rates into the 1 keV flux densities.\\

\begin{table}
\caption{Log of the {\it Chandra} observations of the hotspots. Columns:
 1-hotspot name; 2-{\it Chandra} observation ID; 3-observation date; 4-time on source  after filtering for flaring
background.}
\begin{center}
\begin{tabular}{llccc}
\hline
Name           &ObsID      &Date                 &Livetime                     \\
    &     & (YYYY-MM-DD)    &(ksec)                     \\
\hline
3C\,105 S   &9299       &2007-05-12             &8.1                \\
3C\,195 S   &11501      &2010-01-09             &19.8              \\
3C\,227 E/W &6842       &2006-01-15             &29.8               \\
            &7265       &2006-01-11             &19.9               \\
\hline
\end{tabular}
\end{center}
\label{Xobs}
\end{table}

\subsection{Radio observations}

We retrieved archival VLA data for the hotspots 3C\,105 South, 3C\,227 East
and West, and 3C\,195 South. 
 Observations of 3C 105 South at 4.8 and 8.4 GHz, and 3C 227
West at 8.4 GHz were performed with the array in A-configuration and 
were centred on the hotspot itself, while
observations at 4.8 GHz for 3C 195 and 3C 227 were performed with the 
array in B-configuration and were centred
on the nucleus of the radio galaxy. In all the observations the absolute 
flux density scale was calibrated using the primary calibrator 3C\,286. 
The phase calibrators were 0424$+$020, 0730$-$116, and 0922$+$005 for 
3C\,105, 3C\,195, and 3C\,227, respectively. The observations were 
performed with the historical VLA and the bandwidth was 50 MHz per per intermediate frequency (IF), 
with the exception of the observations of 3C\,105 South and 3C\,227 West that 
had a bandwidth of 25 MHz per IF, and 12.5 MHz per IF, respectively. 
Calibration and data
reduction were carried out following the standard procedures
for the VLA implemented in the National Radio Astronomy
Observatory (NRAO)'s Astronomical Image Processing
System (AIPS) package. Final images were produced after
a few phase-only self-calibration iterations and using uniform weighting 
algorithm. Primary beam
correction was applied at the end of the imaging process.
The rms noise level on the image plane is negligible if
compared to the uncertainty of the flux density due to
amplitude calibration errors that, in this case, are estimated
to be $\sim$ 3 per cent. Log of the radio observations is reported
in Table 3.

In addition to the archival data, we got Jansky VLA observations at 22 
GHz of the hotspots 3C\,227 West and East (project code 18A-087). 
Observations were performed on 2018 March 5 with the array in A-configuration. Details on the observations and data calibration and 
imaging are discussed in \citet{mo20}.

\begin{table*}
\caption{VLA observations of the radio hotspots. Columns: 1-hotspot
  name; 2-frequency; 3 \& 4-beam size and beam
  position angle, respectively; 5-off-source 1$\sigma$ noise level
  measured on the radio image; 6 \& 7-date of the observations
  and project code, respectively; 8: target offset from the pointing direction.}
\begin{center}
\begin{tabular}{lcccclll}
\hline
Name & Freq. & Beam & PA & rms & Date & Code & Offset from the pointing centre \\
     & GHz   & arcsec & deg & mJy/beam &  & \\
\hline
3C\,105 S & 4.8 & 0.38$\times$0.32 & 57 & 0.05& 19-07-2003 & AM772 &on source\\
3C\,105 S & 8.4 & 0.23$\times$0.12 & 38 & 0.05& 19-07-2003 & AM772 &on source\\
3C\,195 S  & 8.4 & 1.30$\times$0.80 &$-$16& 0.11 & 17-01-2004 & AM772 &on source\\
3C\,227 E & 4.8 & 1.25$\times$1.19 & 37 & 0.08 & 13-07-1986 & AS264 &110\arcsec\\
3C\,227 W\tablenotemark& 4.8 & 1.25$\times$1.19 & 37 & 0.08 &13-07-1986 & AS264  &110\arcsec\\
3C\,227 W & 8.4 & 0.38$\times$0.23 & 48 & 0.03 & 25-05-1990 & AB534 &on source\\ 
\hline
\end{tabular}
\end{center}
\label{radio-data}
\end{table*}

\begin{figure*}
\begin{center}
\includegraphics{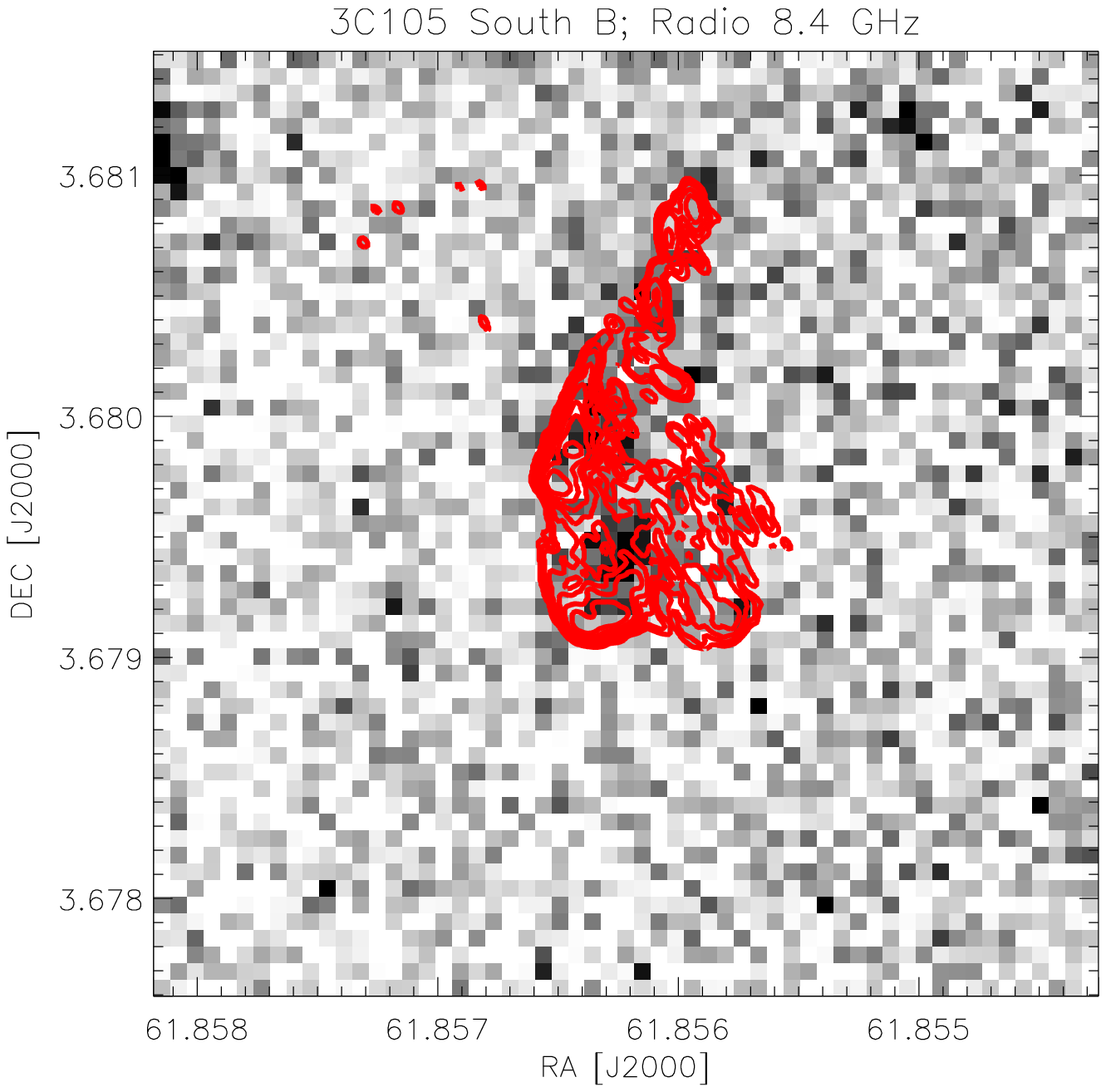}
\includegraphics{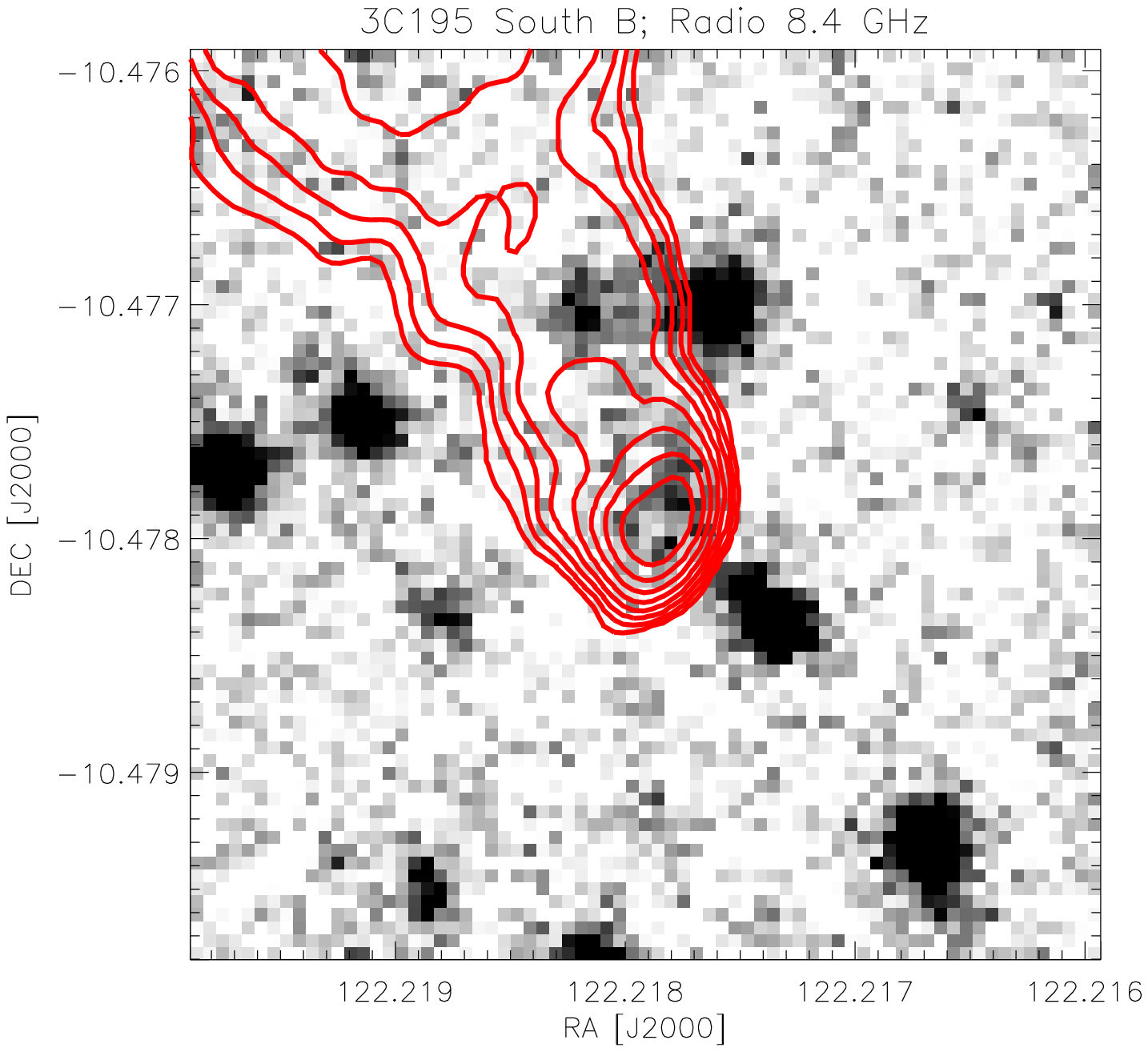}
\includegraphics{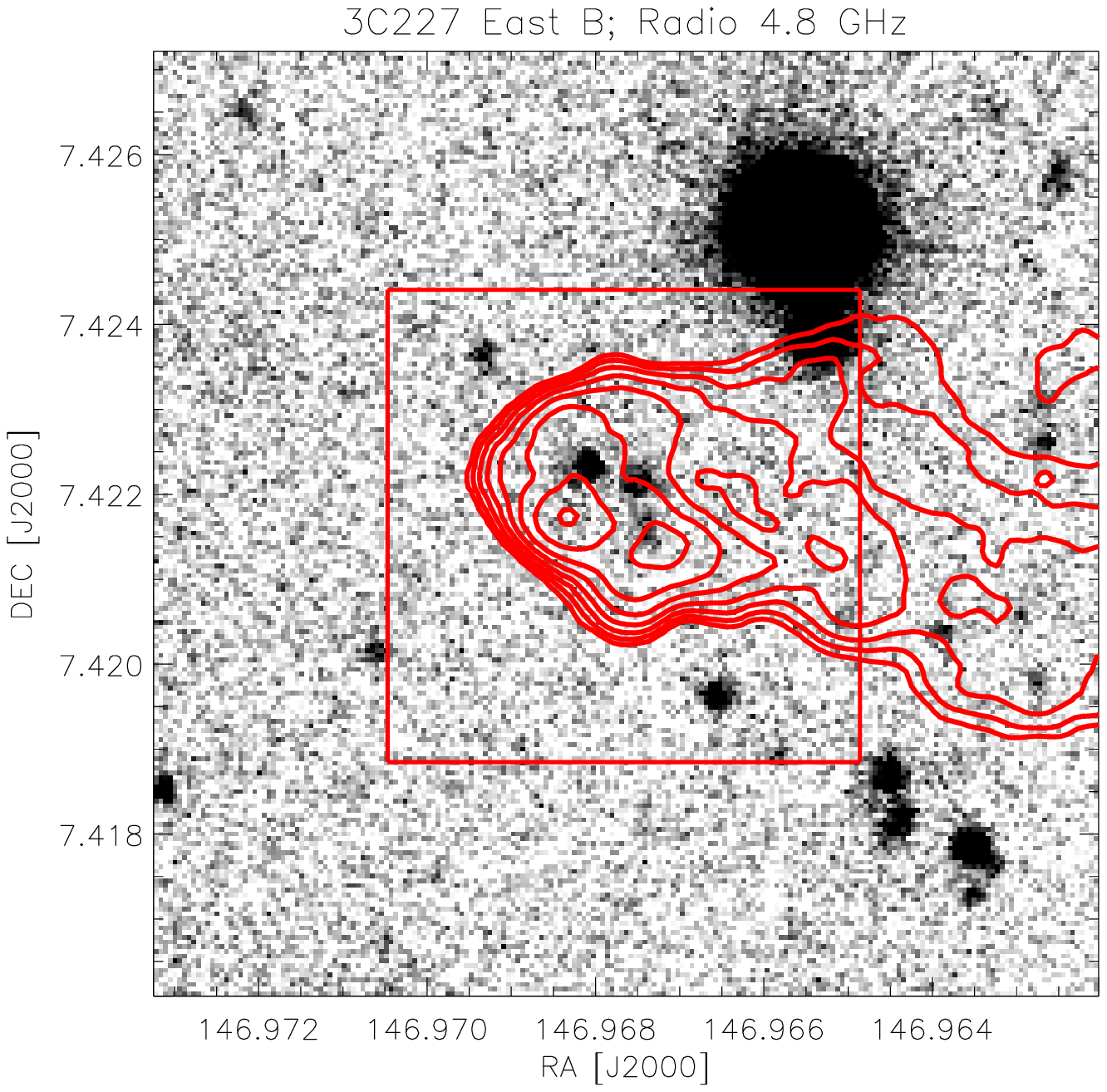}
\includegraphics{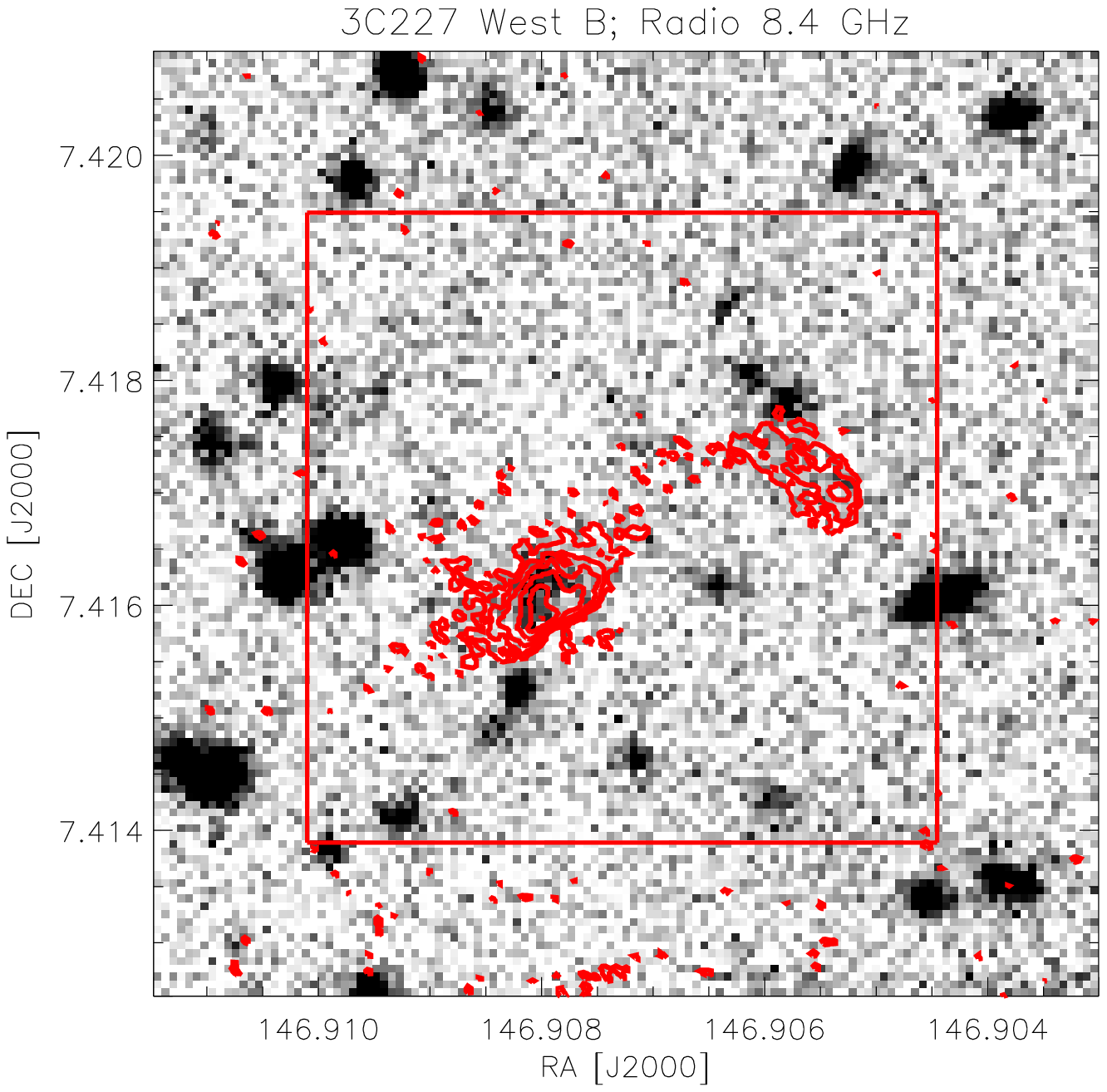}
\vspace{17cm}
\caption{B-band optical images and overlaid radio contours of the
  hotspots 3C\,105 South ({\it top left}), 3C\,195 South ({\it
    top right}), 3C\,227 East ({\it bottom left}), and 3C\,227 West
  ({\it bottom right}). The radio frequency is reported on each panel. The first contour is three times the off-source
  noise measured on the image plane and reported in Table 
  \ref{radio-data}. Contours increase by a factor of 2. In the bottom panels the
  red box represents the area shown in
  Figs. \ref{3c227east_figure} and \ref{3c227west_figure}.}
\label{radio-optical-images}
\end{center}
\end{figure*}

\section{Results}
\subsection{Image registration and flux measurements}

To construct the SED of individual hotspot components, the flux density at the various wavelengths must be measured in the same region, avoiding at the same time
contamination from unrelated sources and components. Radio and optical/NIR images
were aligned with respect to each others 
using reference sources from the 2MASS catalogue (see Sec. 2.1). Figure \ref{radio-optical-images} shows the 
optical B-band image of each hotspot with superimposed radio contours. \\
We performed the astrometric correction of 
the X-ray images by comparing the cores' X-ray and radio positions and
verified the accuracy of the registration (within 
0.1 arcsec) using sources with  infrared counterparts in the
2MASS.\\
We defined a common region of integration for the radio, optical/NIR and X-rays
images, following the contour on the radio emission that corresponds
to the 5 per cent of the radio peak flux of each hotspot component. 

\begin{figure*}
\begin{center}
\includegraphics{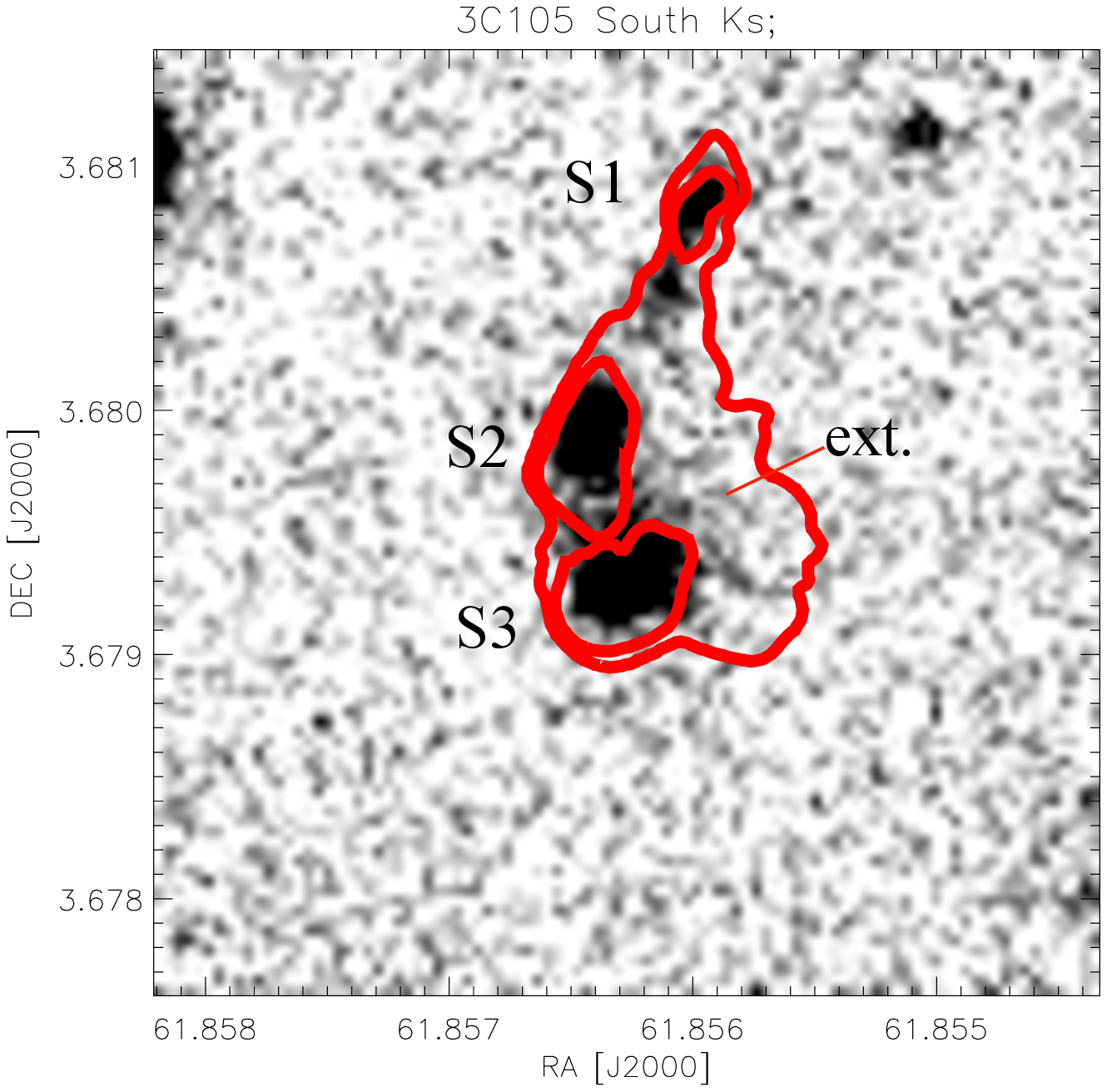}
\includegraphics{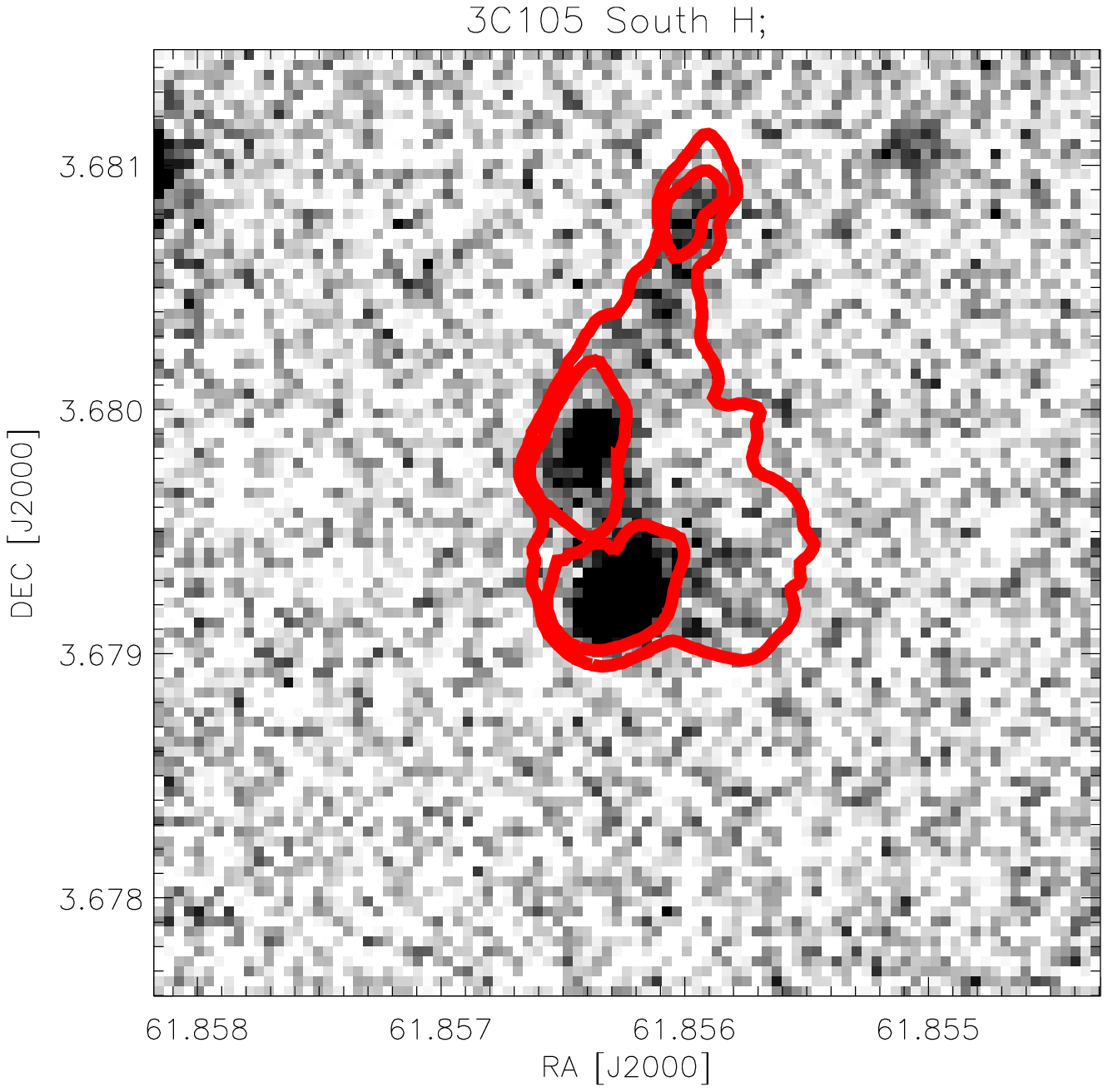}
\includegraphics{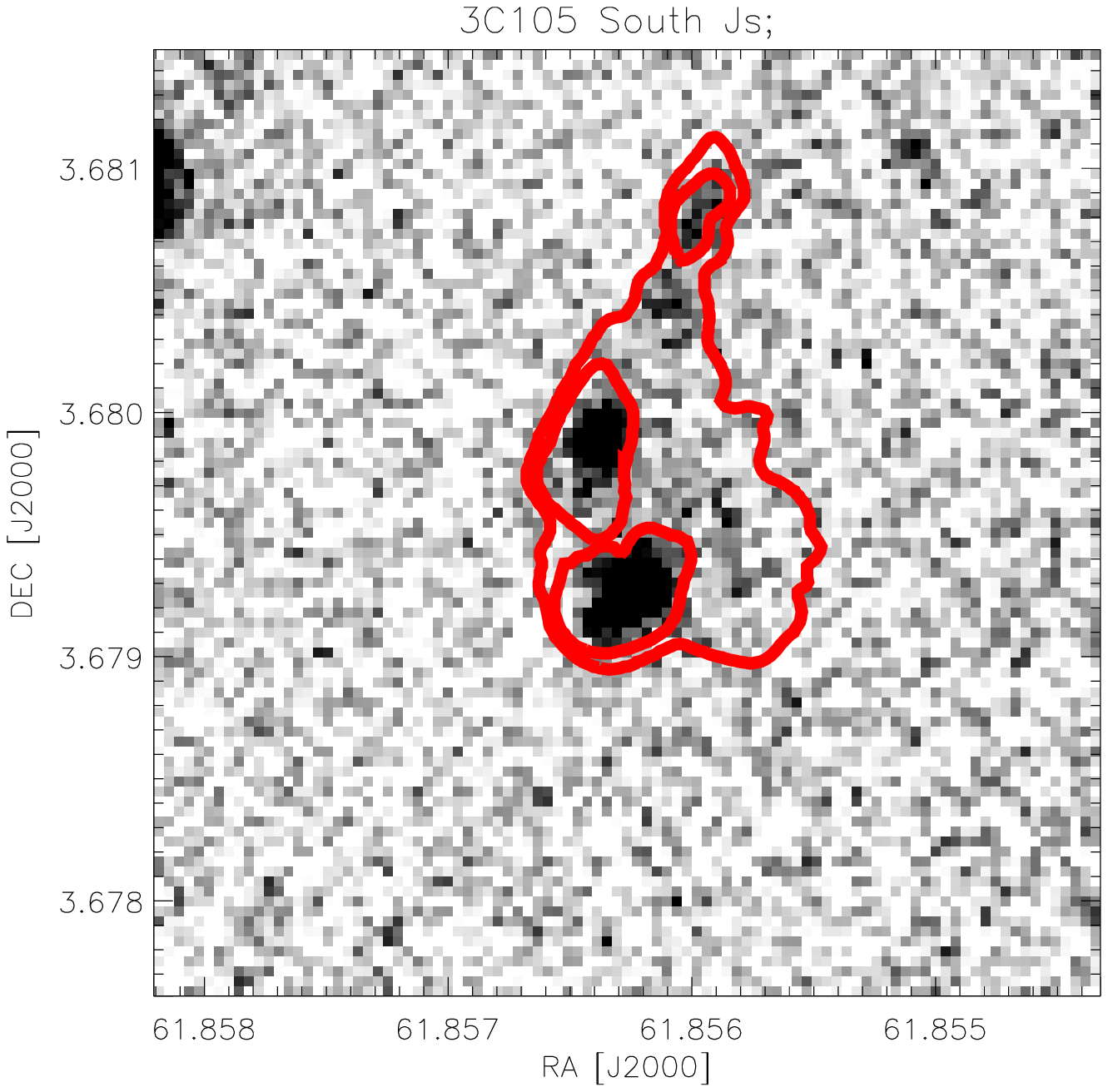}
\includegraphics{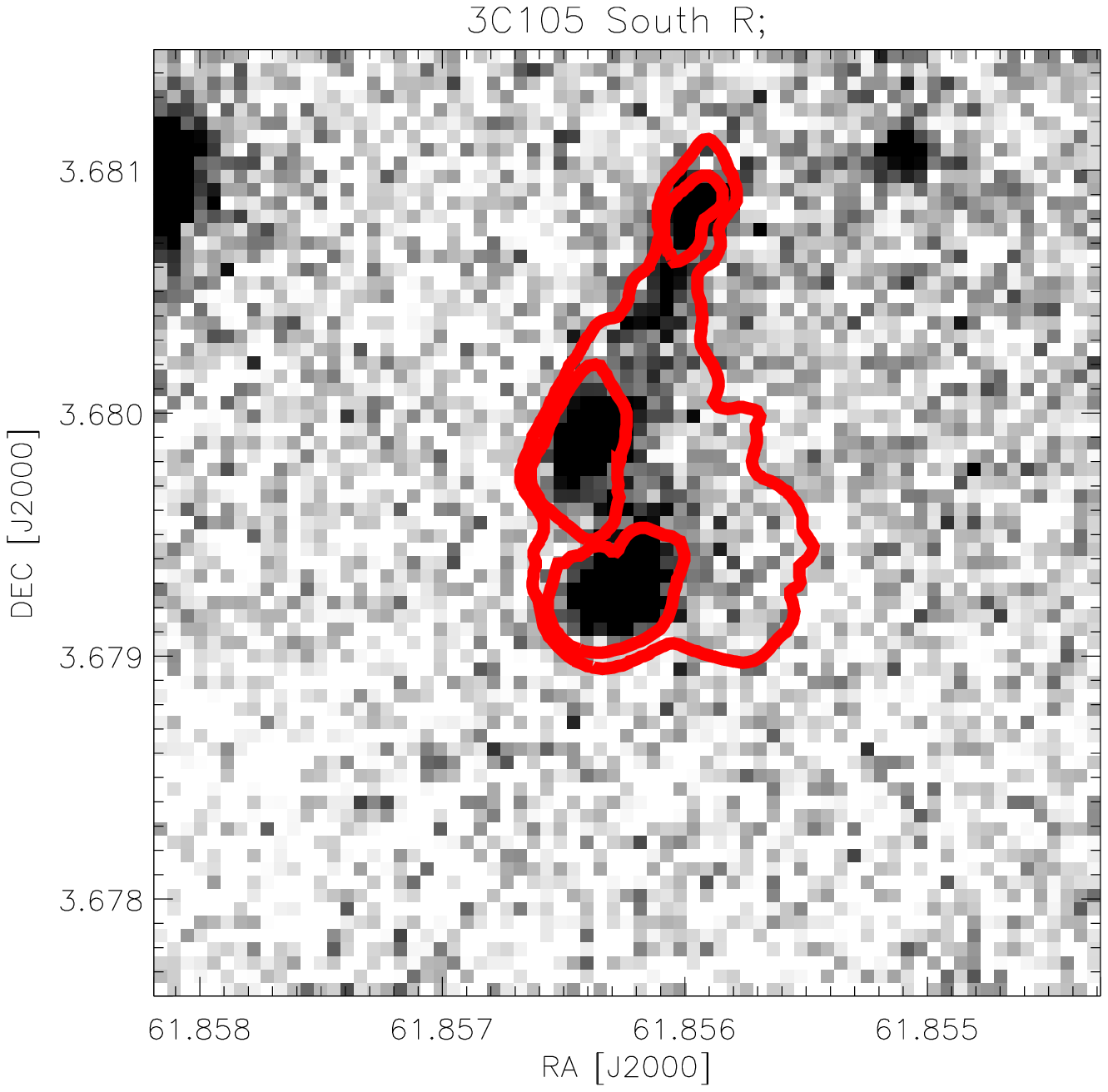}
\includegraphics{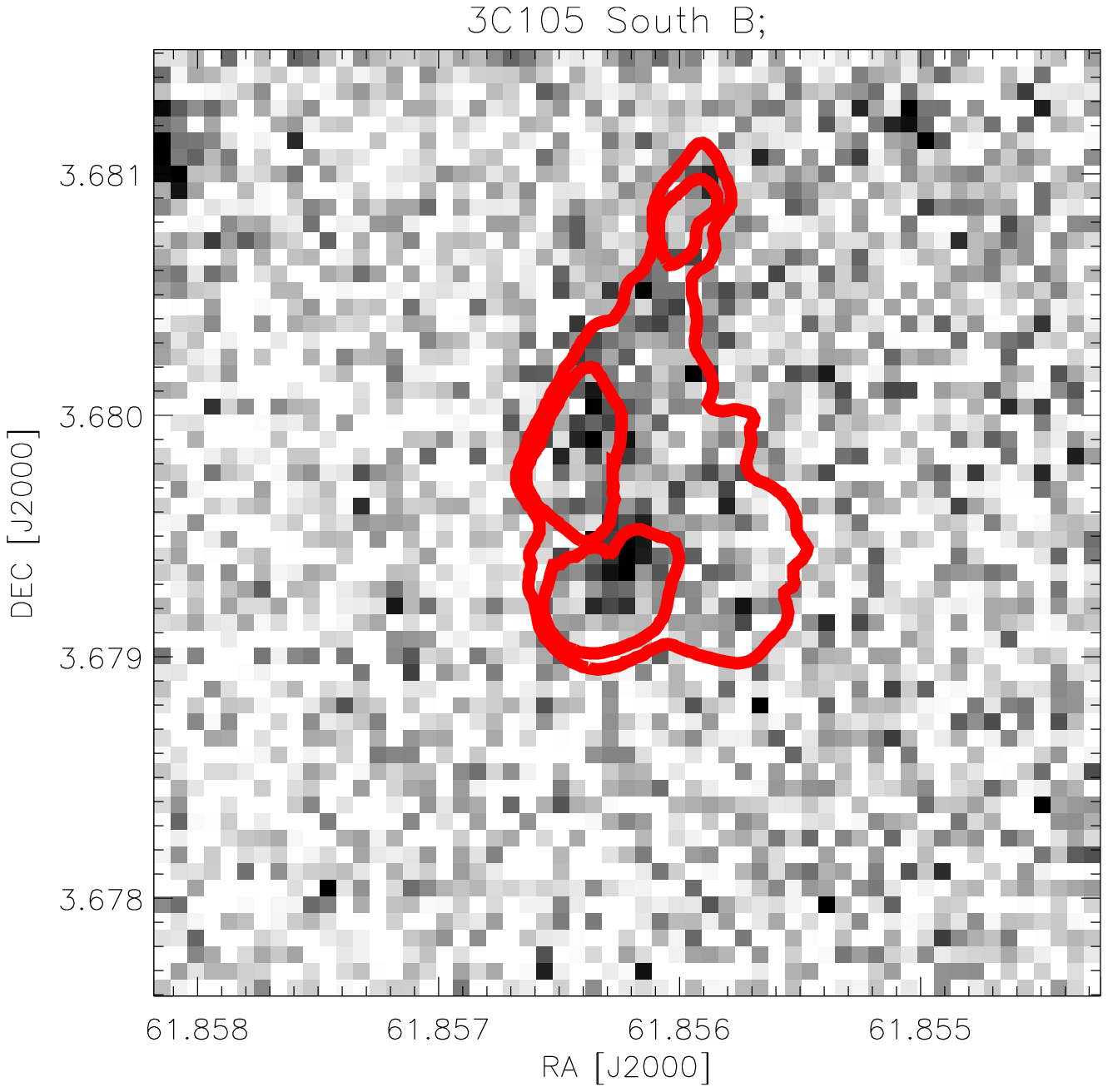}
\includegraphics{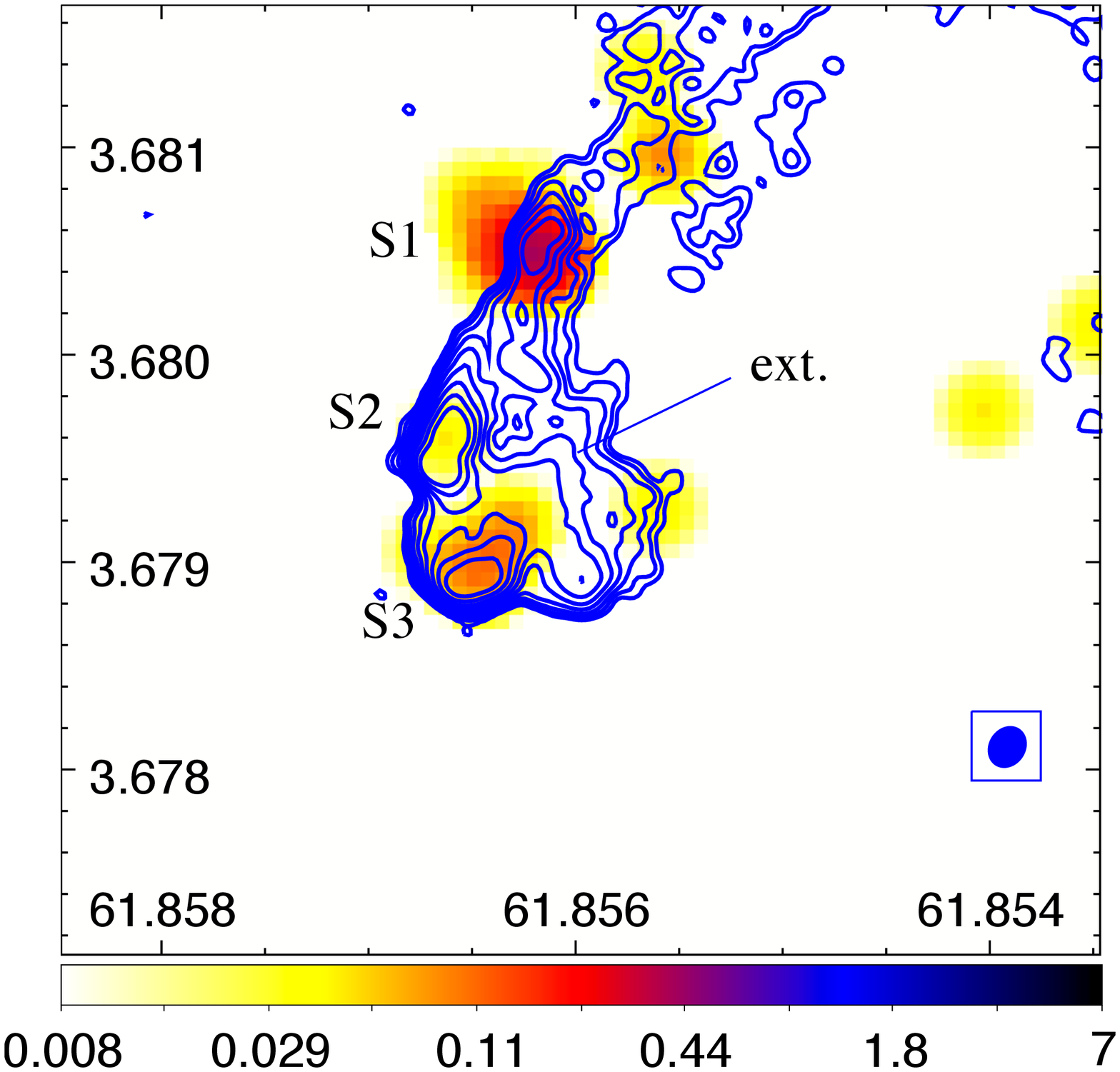}
\vspace{12cm}
\caption{Multi-band images of 3C\,105 South. From left to right
  and top to bottom: VLT K s, H, J s, R, and B bands, and {\it Chandra} X-ray
  images. In the NIR/optical images the first contour represents three
times the off-source noise measured on the image at 8.4 GHz and reported in Table
\ref{radio-data}, while the other contours are the extraction regions
considered for the flux measurement of the three compact components (S1, S2, S3) and correspond to the 5 per cent of
the radio peak flux of each component.
The intensity scale of the 0.5--7 keV image (in count/pixel units) is logarithmic and the image was binned to half the
native pixel size (0.246 arcsec) and smoothed with a
Gaussian using $\sigma$ = 4.
The radio contours at 4.8 GHz are overlaid on the X-ray image.  The
first radio contour represents three
times the off-source noise measured on the image and reported in Table
\ref{radio-data},  while dashed contours indicate the first negative contour ($-$3$\times$rms). Contours increase by a factor of 2. The radio beam is shown in the box in the bottom right corner of the X-ray image.}
\label{3c105_figure}
\end{center}
\end{figure*}

NIR and optical images with the regions of
integration and {\it Chandra} X-ray
images with overlaid radio contours are presented in Figs. \ref{3c105_figure} to \ref{3c227west_figure}. 
Radio, NIR and optical fluxes for each hotspot
component are reported in Table \ref{fluxes}.
There are discrepancies in some bands between our measurements and those reported by \citet{mack09} and \citet{mo12} for 3C 105 South and 3C195 South. The reason of the discrepancy boils down to few facts: (a) in this work we have performed a local evaluation of the sky-background, as the sky residuals in the background-subtracted product of the pipeline were not negligible; (b) we have adopted polygonal region for flux integration following a fixed iso-contour level, which is different from what used in the past; and (c) for the exposures that have no available standard star observations on the same night, we used field stars to calibrate them, whereas the previous studies used standard star observations acquired on different nights, with probably different atmospheric conditions.\\
\indent
 For 3C\,227 West, the X-ray counts were enough to obtain spectra of each
of the two components of the hotspot complex.
An absorbed power-law model with the column density fixed to the
Galactic value \citep[$\rm N_H=2\times10^{20}$ cm$^{-2}$][]{nhref} was used to
simultaneously fit the spectra of the two {\it Chandra} observations. The best
fit values for each spectrum are reported in Table \ref{Xrayspec}.  
The 0.5--7.0 keV net count rates of 3C\,105 South, 3C\,195 South and
3C\,227 East were converted into unabsorbed flux densities at 1 keV
assuming an absorbed power law model with photon index $\Gamma=$ 1.8 (in accordance with the best fit model for 3C\,227 West)  and the column
density fixed to the Galactic value (Table \ref{Xrayspec}).
Keeping into account differences in the selected regions, our results
are in broad agreement with those reported for the targets in the
literature \citep{hardcastle07,massaro11,mo12,mingo17}. \\

\begin{table*}
\caption{Radio, NIR and optical flux densities of the hotspot components. }
\begin{center}
\begin{tabular}{lcccccccc}
\hline
Name & 4.8 GHz & 8.4 GHz &  Ks    &   H   &    Js  &   R   & B      \\ 
     &(mJy)     & (mJy)     &($\mu$Jy) &($\mu$Jy) &($\mu$Jy) &($\mu$Jy) &($\mu$Jy)
\\ 
\hline
3C\,105 S1& 26.4$\pm$0.7&18.4$\pm$0.5&2.6$\pm0.2$& 2.8$\pm0.2$
&1.1$^{+0.2}_{-0.2}$&0.3$^{+0.4}_{-0.2}$& $<$0.10 \\ 
3C\,105
S2&540$\pm$16&372$\pm$11&14.4$\pm0.4$&13.5$^{+0.7}_{-0.6}$&5.8$^{+1.0}_{-0.9}$&1.0$^{+1.2}_{-0.5}$&0.2$\pm$0.1
\\ 
3C\,105
S3&403$\pm$12&260$\pm$8&23.4$\pm$0.5&22.7$\pm$0.9&8.5$^{+1.5}_{-1.3}$&1.3$^{+1.6}_{-0.7}$&0.3$\pm$0.1\\ 
3C\,105 S
Ext&275$\pm$8&130$\pm$4 &17$\pm2$ &17$\pm$2 &6.4$^{+4}_{-3}$ 
&1.4$^{+5.0}_{-1.4}$ &0.9$^{+0.5}_{-0.3}$ \\
3C\,195 S& - &94$\pm$3& 3.3$^{+0.8}_{-0.7}$&$<$0.46& -
&0.26$^{+0.01}_{-0.02}$&0.14$^{+0.01}_{-0.01}$ \\ 
3C\,227 E1&102$\pm$3& - &$<$1.10 & - & - & 0.19$^{+0.07}_{-0.06}$ & $<$0.14 & \\
3C\,227 E2&84$\pm$3& -  &$<$1.10 & - & - &0.4$^{+0.1}_{-0.1}$&0.5$^{+0.2}_{-0.1}$ \\ 
3C\,227 W1&90$\pm$3&63$\pm$2&11$^{+2}_{-1}$&9.5$^{+0.6}_{-1.3}$&5.3$^{+0.7}_{-0.6}$&1.2$\pm0.1$&1.0$\pm$0.2\\ 
3C\,227 W2&28.3$\pm$0.8&15.7$\pm$0.5&9.5$\pm$1.3&4.3$\pm$0.8&2.9$\pm$0.4&0.44$^{+0.07}_{-0.06}$&0.4$\pm$0.1\\ 
\hline
\end{tabular}
\end{center}
\label{fluxes}
\end{table*}

\begin{table*}
\caption{X-ray fluxes of the hotspot components and knots. An absorbed
  power-law model was assumed to estimate the fluxes. Columns: 1-radio component; 2-X-ray photon index, (f)=fixed; 3-Galactic
  column density in cm$^{-2}$; 4-unabsorbed flux densities at 1 keV. Uncertainties are reported at 90\% confidence level. (*): 3$\sigma$ upper limit. }
\begin{center}
\begin{tabular}{lccc}
\hline
Component                                       &$\Gamma$   &N$_{H,Gal}$       &F$_{1keV}$\\
                                                &           &cm$^{-2}$         &erg cm$^{-2}$ s$^{-1}$\\

\hline
3C\,105 S1      &1.8(f)     &10.4$\times$10$^{20}$    &(5.2$\pm$1.3)$\times$10$^{-15}$\\
3C\,105 S2+S3   &1.8(f)     &''                       &(2.1$\pm$0.8)$\times$10$^{-15}$\\
3C\,105 S Ext   &1.8(f)     &''                       &$<$0.9$\times$10$^{-15}$(*)\\
3C\,195 S       &1.8(f)     &7.8$\times$10$^{20}$     &(1.3$\pm$0.4)$\times$10$^{-15}$\\
3C\,227 E       &1.8(f)       &2$\times$10$^{20}$       &(1.0$\pm$ 0.2)$\times$10$^{-15}$\\                        
3C\,227 W1      &1.8$\pm$0.3  &''                      &(4.0$\pm$0.7)$\times$10$^{-15}$\\
3C\,227 W2      &1.8$\pm$0.5  &''                      &(2.1$\pm$0.7)$\times$10$^{-15}$\\
\hline
\end{tabular}
\end{center}
\label{Xrayspec}
\end{table*}

\subsection{Notes on individual sources}

\subsubsection{3C\,105 South}
The hotspot complex consists of a jet knot (S1), observed from radio to X-rays,
 and a double hotspot (see Fig. \ref{3c105_figure}). 
The detection of NIR emission from 3C\,105 South was first reported in \citet{mack09}. \citet{mo12} presented a radio-to-X-ray study of the compact features of the hotspot complex, while here we focused on the diffuse component.
The primary hotspot (S2)
is brighter in radio than the secondary hotspot (S3), while it
becomes the faintest in NIR and optical (Fig. \ref{3c105_figure}),
suggesting that the energy distribution of the radiating particles in the two hotspots is different. 
In the {\it Chandra} observations the jet knot S1 is clearly observed, while X-ray emission from S2 and S3 is only marginally detected (6.6$\pm$2.6 net counts in the 0.5--7 keV energy range).

\begin{figure}
\begin{center}
\includegraphics{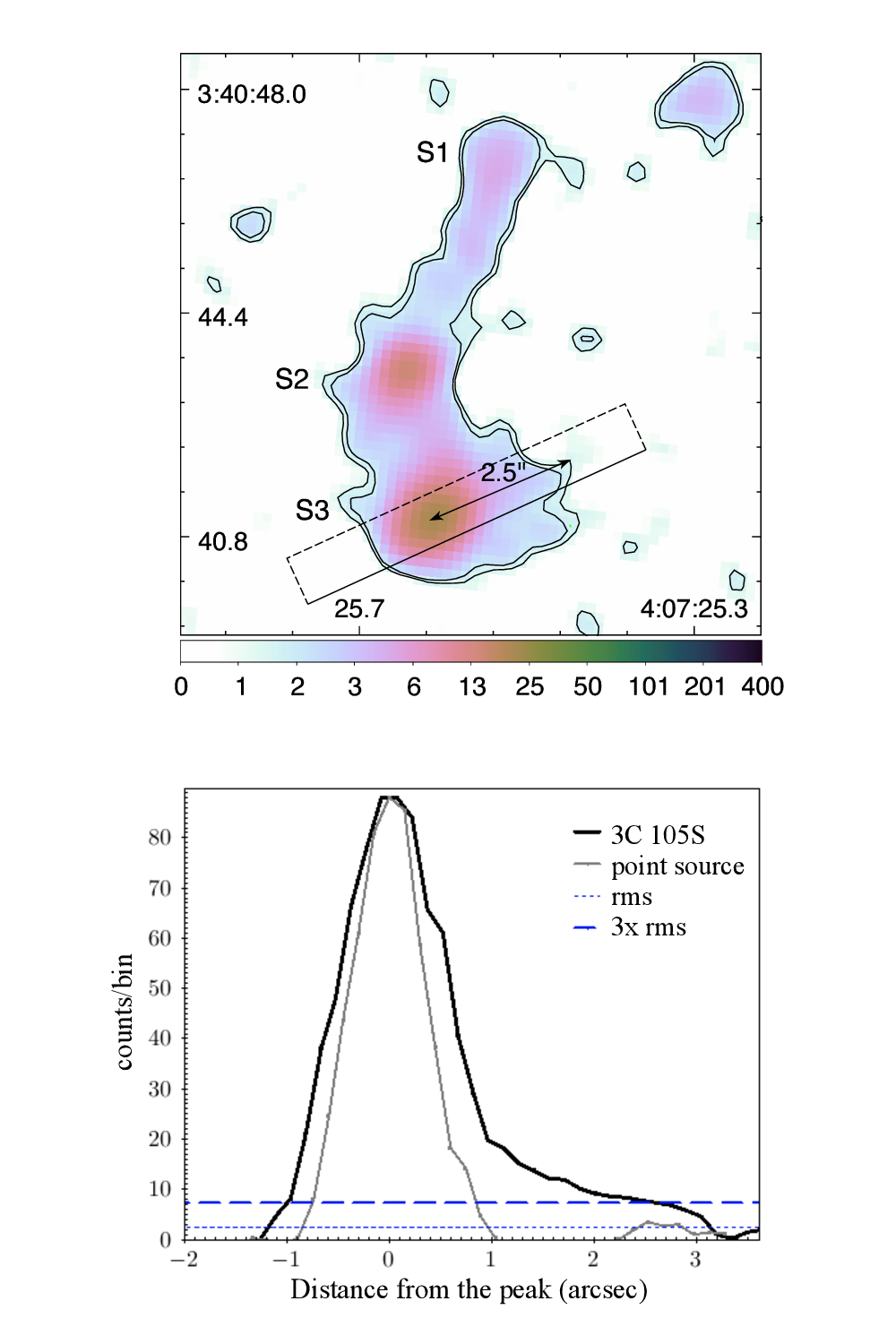}
\vspace{15.5cm}
\caption{ Top: H band image of 3C\,105 South. The image was smoothed with a Gaussian using $\sigma=$3. The contours start at 3$\times$rms and increase by a factor of $\sqrt{2}$. The box corresponds to the extraction region of the brightness profile. Bottom: H band brightness profile of the S3 and downstream region (black thick line). The profile of a nearby point-like source in the field is also shown (thin solid line). The background level is given by the dashed lines.}
\label{3c105profile}
\end{center}
\end{figure}

The three compact components of the hotspot complex are enshrouded by diffuse emission (3C~105 S ext) that could not be well characterized in \citet{mo12}.
The NIR/optical diffuse emission has been obtained by subtracting the emission of the three main components from the total flux: with the new J- and B-band data, the diffuse emission is now detected in all radio-to-optical bands (see Table \ref{fluxes}). 
 The extraction regions of the NIR/optical compact emission were defined based on the radio images, with the limit being (conservatively) fixed at 5 per cent of the radio peak of each component. While this choice was dictated by the need to have common regions for the SED, it is still reasonable as long as the radio and NIR/optical peaks of the three features are spatially coincident. The extension of the diffuse emission cannot be easily determined because of its irregular spatial distribution. To get an indicative estimate, we extracted the brightness profile of the H-band emission from a rectangular region (6\arcsec$\times$0\arcsec.8) covering the S3 component and its dowstream region. The profile is shown in Fig. \ref{3c105profile} together with the level of the background and with the profile of a point-like source in the field, rescaled to the peak of S3. Emission above 3$\times$rms level is detected up to 2\arcsec.5 ($\sim$4 kpc) from the peak. We adopted this estimate as a reference value for the projected size of the diffuse emission.\\
\indent
 No significant
X-ray diffuse emission is observed in the hotspot complex. A 3$\sigma$ upper limit was derived from the counts in the total region excluding the three compact components (Table \ref{Xrayspec}).

\begin{figure*}
\begin{center}
\includegraphics{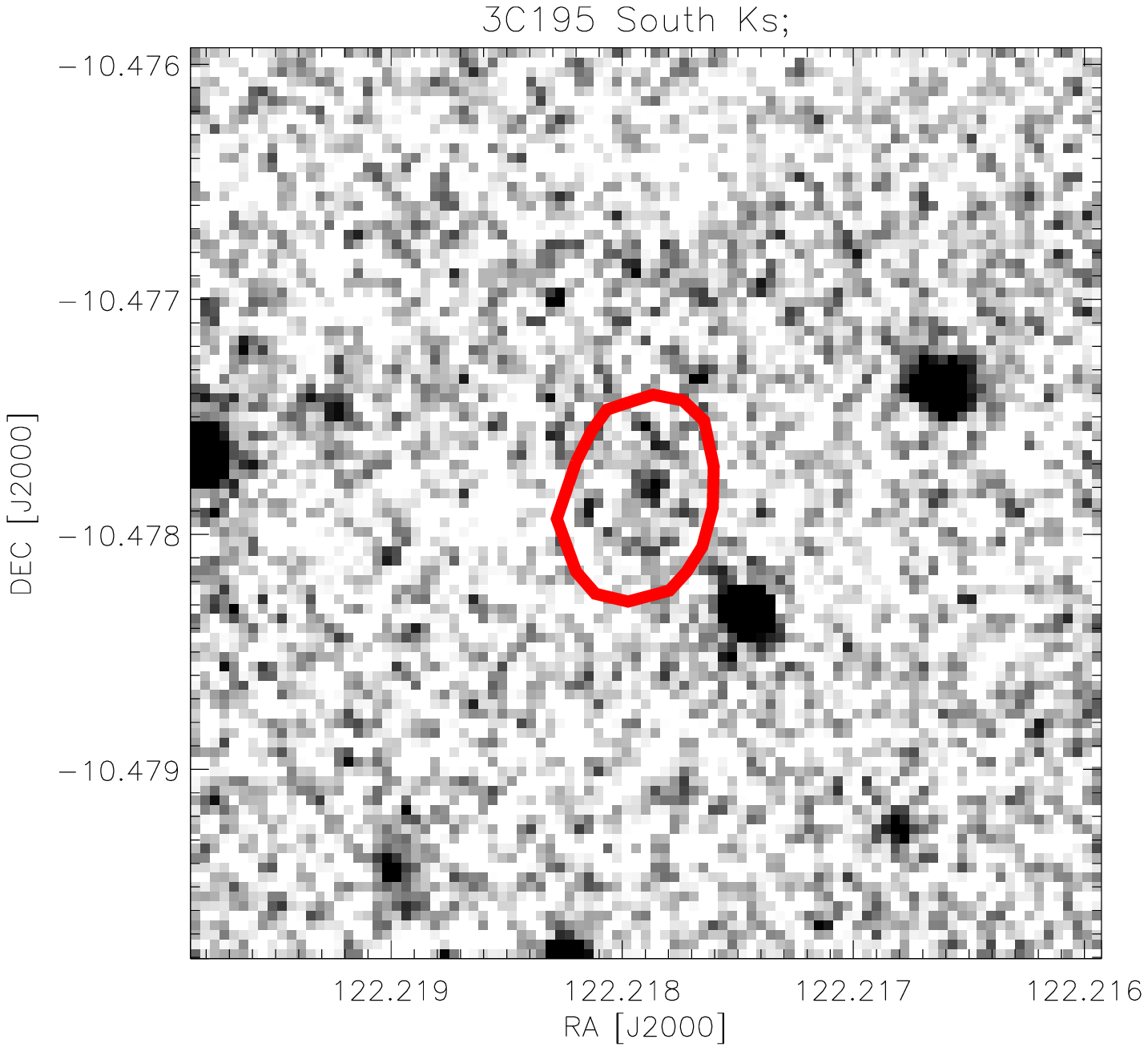}
\includegraphics{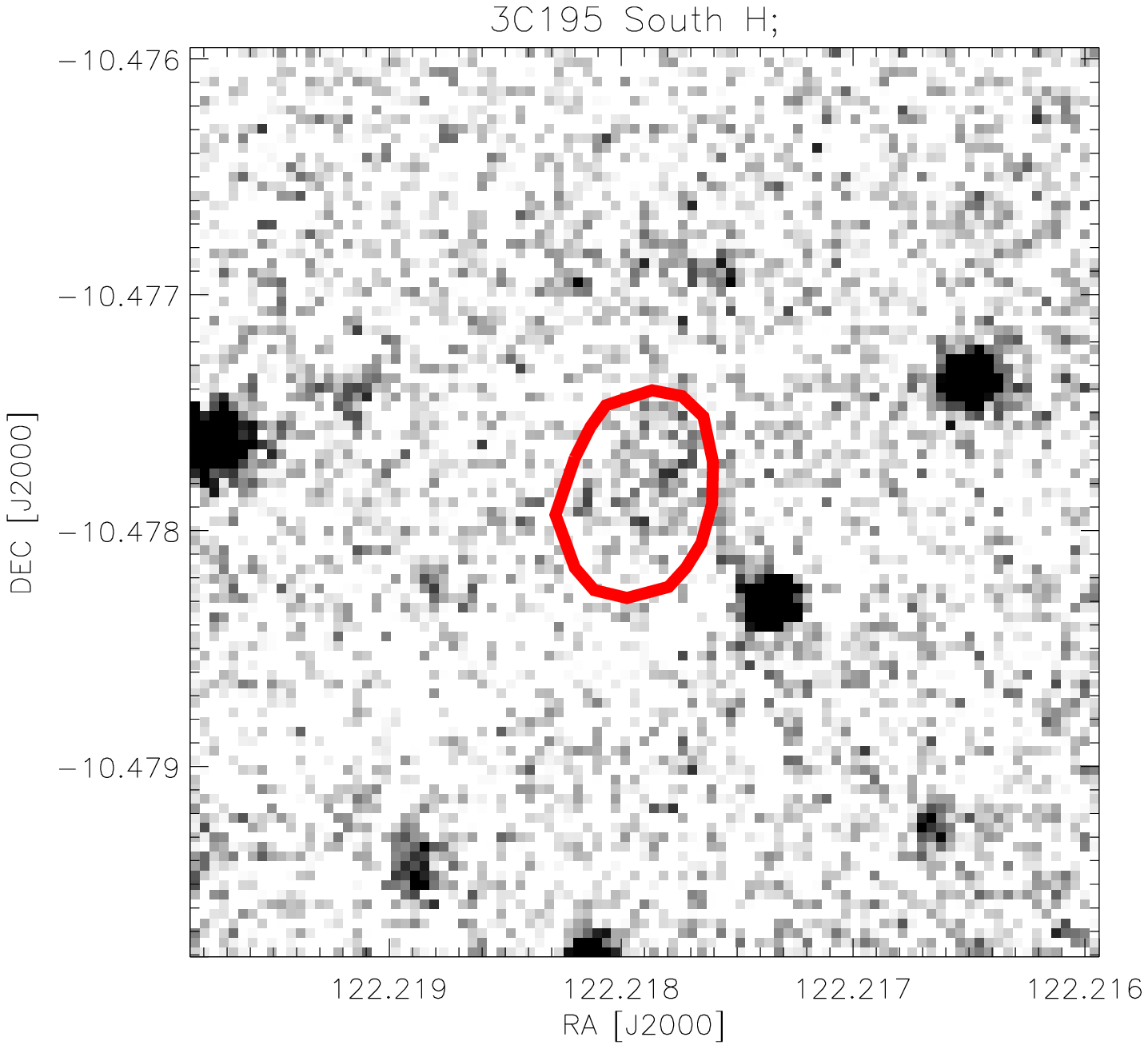}
\includegraphics{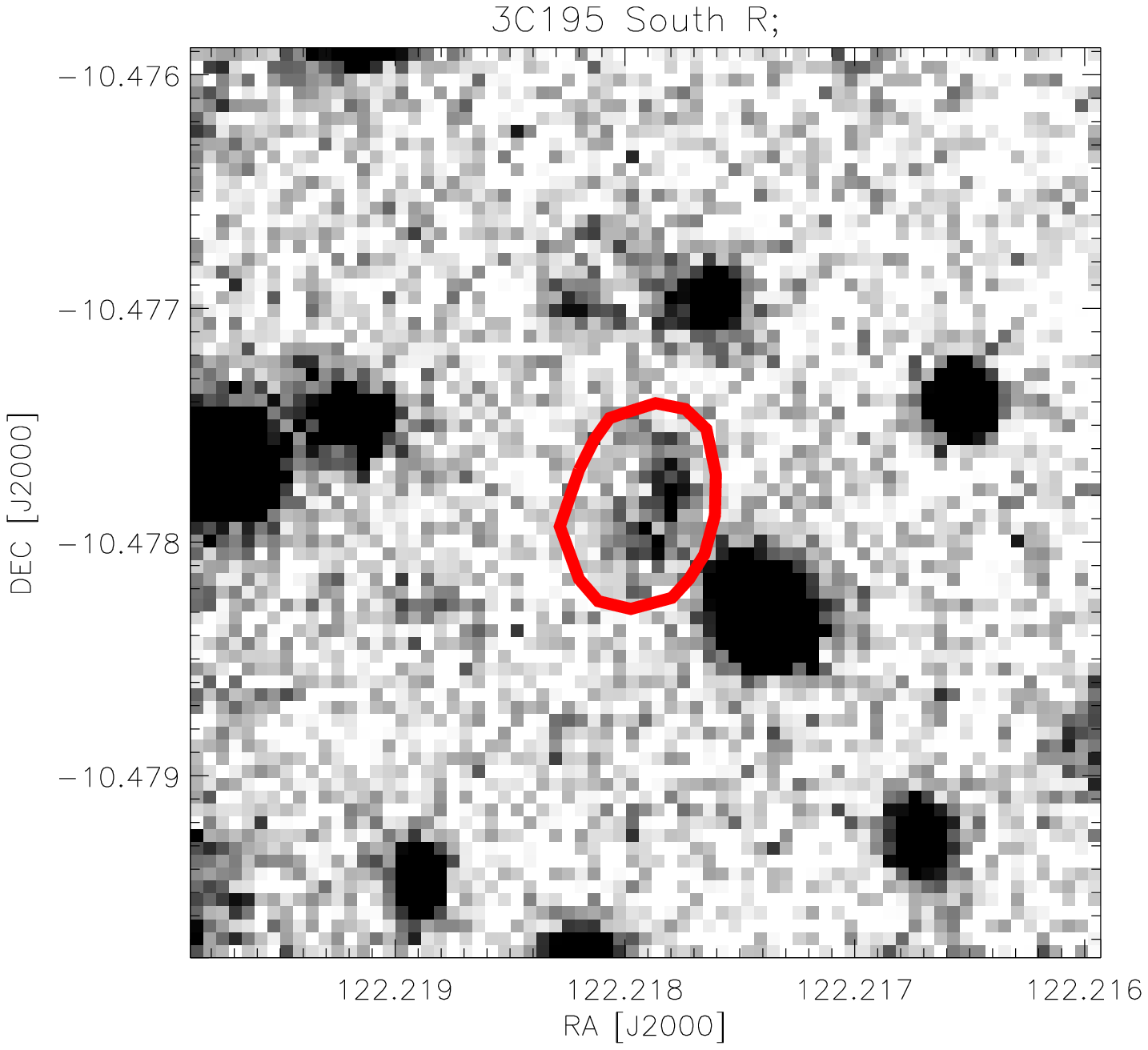}
\includegraphics{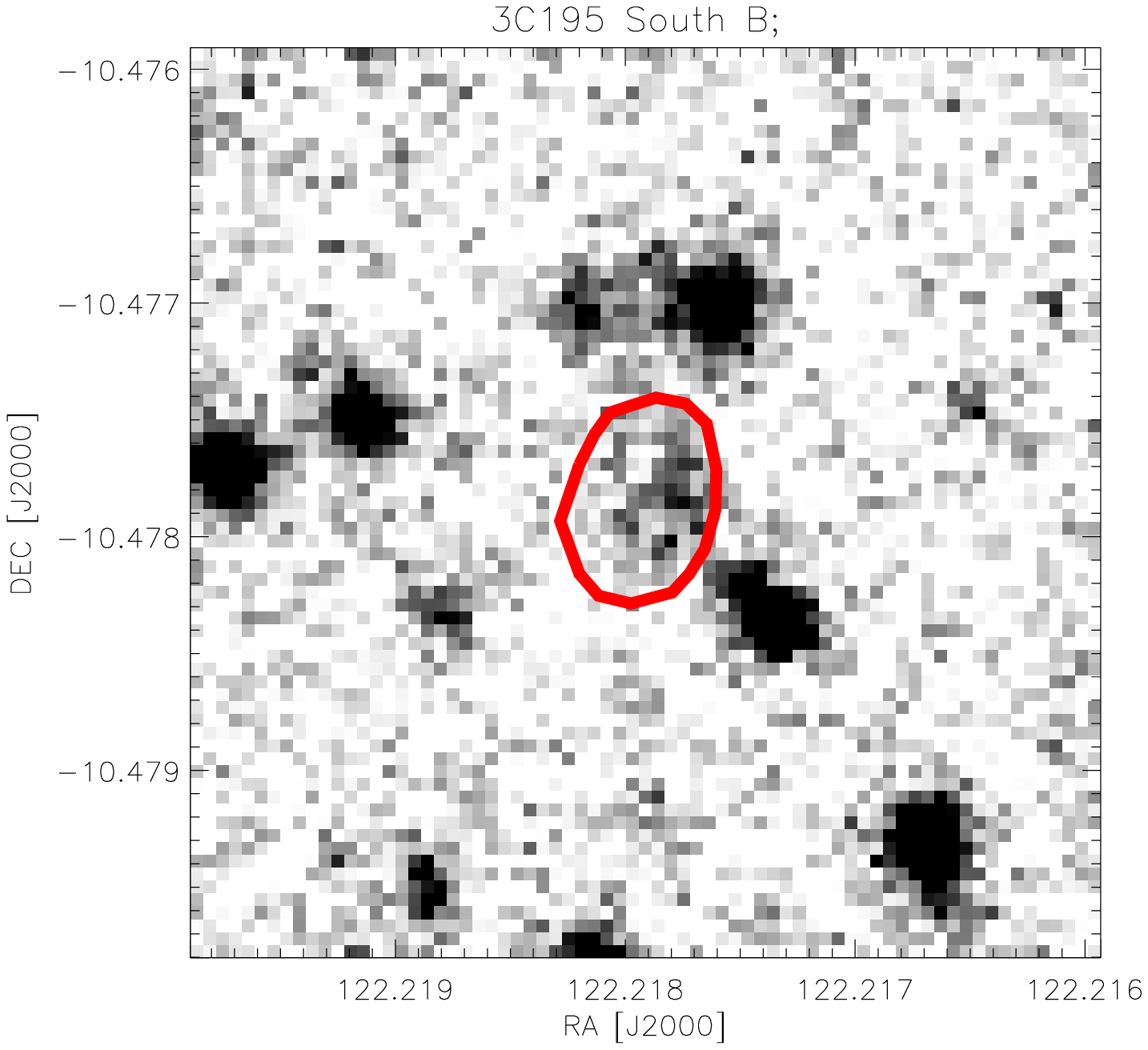}
\includegraphics{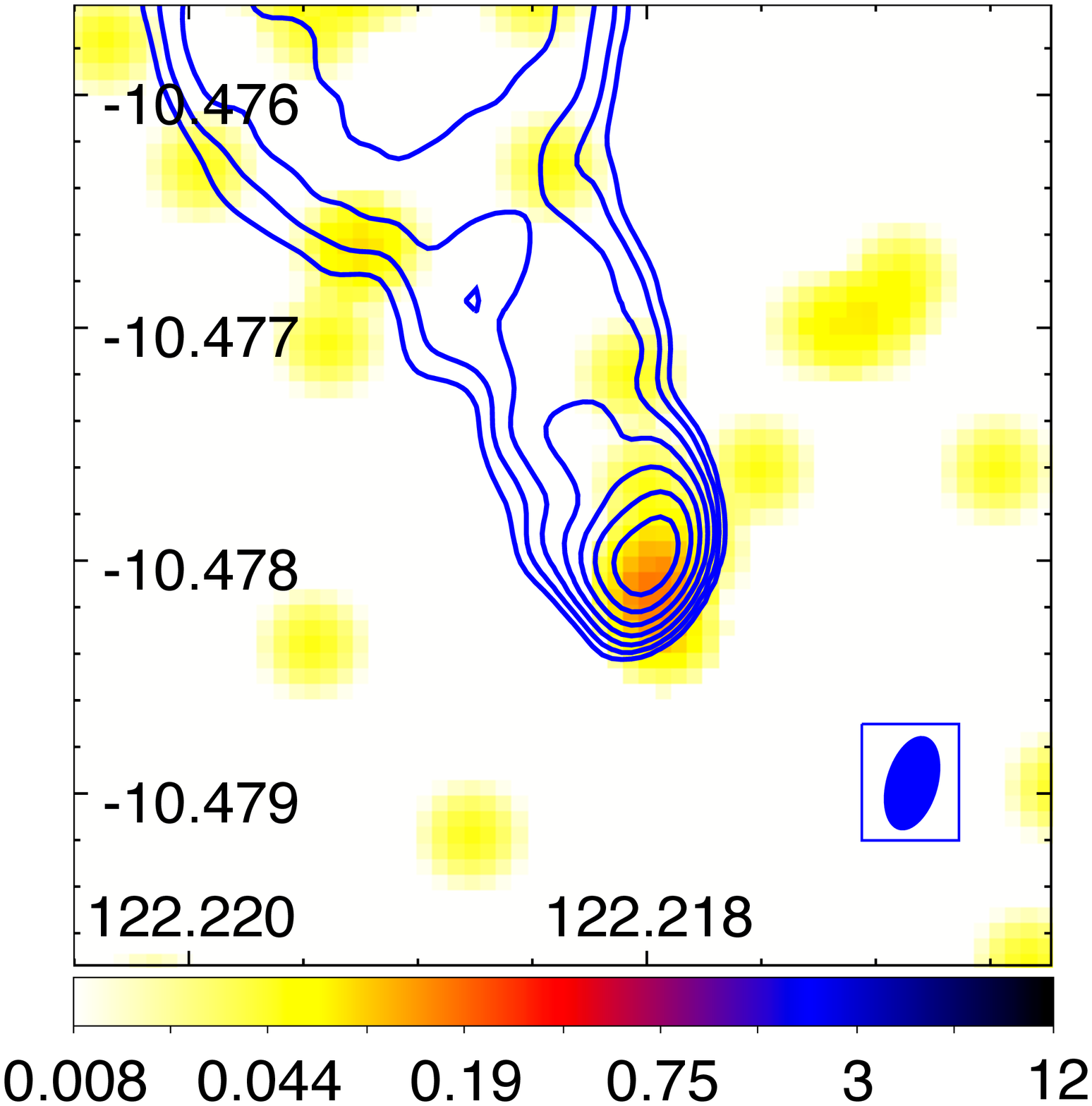}
\vspace{12cm}
\caption{Multi-band images of 3C\,195 South. From left to right
  and top to bottom: VLT Ks, H, R, and B band, and {\it Chandra} X-ray
  images.  In the NIR/optical images the thick red contour represents the extraction region
considered for flux measurement and corresponds to the 5 per cent of
radio peak flux of the hotspot component.  The X-ray
intensity scale (in count/pixel units) is logarithmic and the image was binned to half the
native pixel size (0.246 arcsec) and smoothed with a
Gaussian using $\sigma$ = 4. The radio contours at 8.4 GHz are overlaid on the 0.5$-$7 keV image.
The first radio contour represents three
times the off-source noise measured on the image and reported in Table
\ref{radio-data}, while dashed contours indicate the first negative contour ($-$3$\times$rms). Contours increase by a factor of 2. The radio beam is shown in the box in the bottom right corner of the X-ray image.}
\label{3c195south_figure}
\end{center}
\end{figure*}

\subsubsection{3C\,195 South}
Tentative detection of NIR emission from 3C\,195 South was reported by
\citet{mack09}. The new VLT observations confirm the emission from
this hotspot in K band, while only an upper limit is obtained
in H band. In the optical window, this hotspot is
clearly detected in R- and B-band and, within the extraction region, the emission is extended (Fig. \ref{3c195south_figure}).\\ 
In X-rays, weak, compact emission is detected by {\it Chandra} at $\sim$3$\sigma$ level, in agreement
with the results obtained by \citet{mingo17}. Because of the faintness of the X-ray flux, we cannot be conclusive on the slight offset between the radio and X-ray centroids (Fig. \ref{3c195south_figure}). There is no evidence of diffuse X-ray emission around the compact component.

\begin{figure*}
\begin{center}
\includegraphics{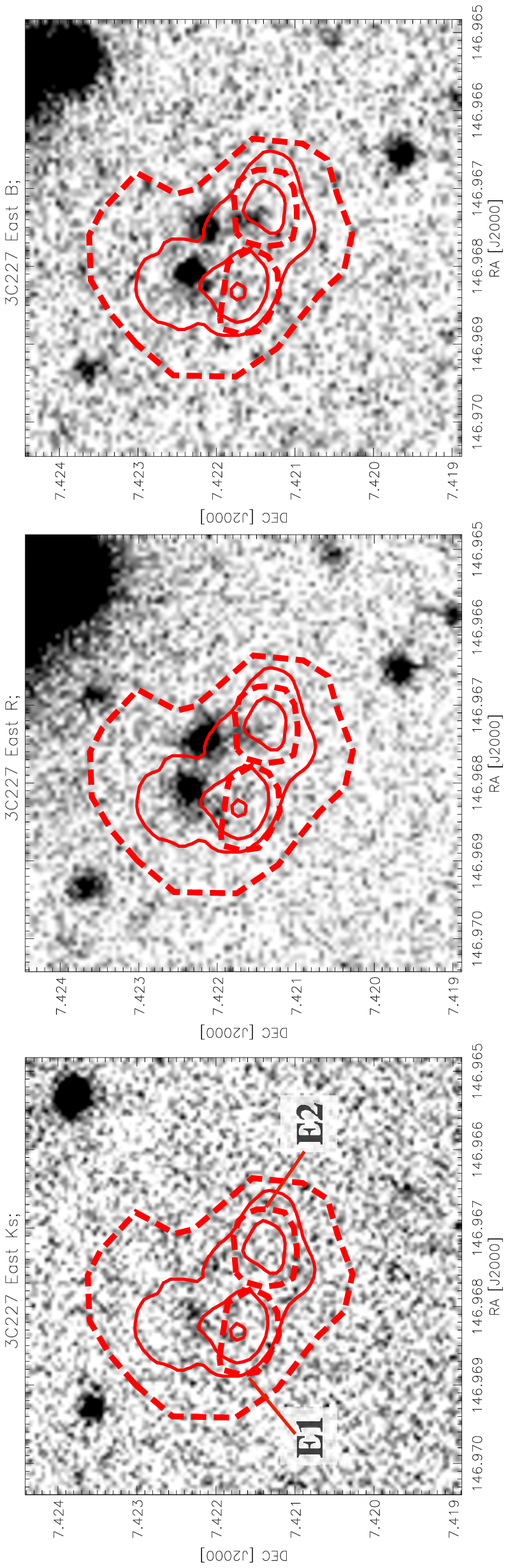}
\vspace{6cm}
\caption{Multi-band images of 3C\,227 East. From left to right:
VLT Ks, R, and B band images. The  thick dashed  red contours are the extraction regions considered for flux measurement. The sub-regions labelled E1 and E2 were selected in order to avoid contamination from foreground stars. The thin radio contours represent 96, 192, and 384 times the off-source noise in Table \ref{radio-data}.}
\label{3c227east_figure}
\end{center}
\end{figure*}

\begin{figure}
\begin{center}
\includegraphics{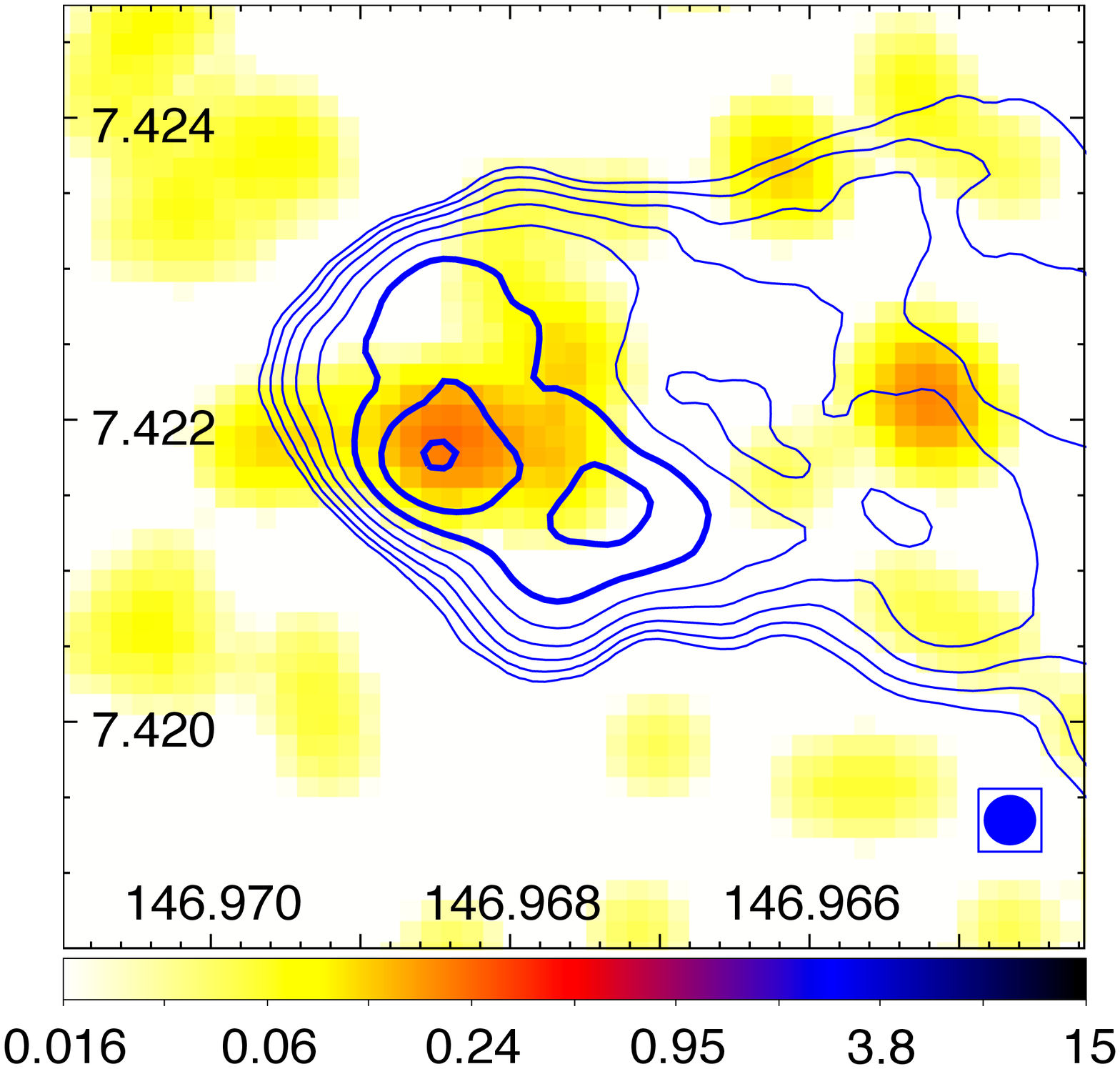}
\vspace{7.8cm}
\caption{{\it Chandra} X-ray
image of 3C\,227 East and overlaid radio contours at 4.8 GHz. The 0.5$-$7
keV X-ray
intensity scale (in count/pixel units) is logarithmic and the image was binned to the native
pixel size (0.492 arcsec) and smoothed with a Gaussian using $\sigma$ =
4. The first radio contour represents three
times the off-source noise measured on the image and reported in Table
\ref{radio-data}, contours increase by a factor of 2.  The thick lines represent the 96, 192 ,384 times the off-source noise as in Figure \ref{3c227east_figure}.  The radio beam is shown in the box in the bottom right corner.}
\label{3c227east_Xray}
\end{center}
\end{figure}

\subsubsection{3C\,227 East}
Emission from 3C\,227 East is clearly seen in the optical R- and B-band \citep[see Fig. \ref{3c227east_figure} and][]{mack09}, while with the new VLT pointing we only measured an upper limit to the K-band flux.
The presence of foreground stars 
in the hotspot complex hampers an accurate determination of the
optical flux for the hotspot components. In order to avoid flux contamination from these stars, we selected two extraction sub-regions (labelled E1 and E2,  see Fig \ref{3c227east_figure}). The fluxes are reported in Table \ref{fluxes}. 
 We checked and no galaxy is reported within E1 and E2 in the Sloan Digital Sky Survey \citep[SDSS,][]{sdss12}, which, in this field, detected galaxies with similar, or fainter, optical fluxes than ours.
We then estimated the probability that the optical emission in the two regions is due to unassociated background galaxies. Using the galaxy number-apparent magnitude relation derived from deep optical surveys \citep[see e.g.][]{madau00}, the expected number of galaxies within each area (approximated to a circle of $\sim$1\arcsec.3 radius) at the measured R and B fluxes is $\sim$0.008 and the probability of having one galaxy within E1 or E2 is $<$1 per cent. 
In addition, extended NIR emission, possibly spatially overlapping with the faint X-ray flux (see below), was reported by \citet[][]{mack09} based on previous VLT observations in 2002, giving further support to a non-thermal origin of the NIR-optical component.
The presence of foreground stars 
precludes also the detection of possible extended optical emission
enshrouding the main hotspot components.\\
{\it Chandra} observations could clearly detect X-ray emission from
the hotspot (Fig. \ref{3c227east_Xray} and Orienti et al. 2020), in agreement with previous
work by \citet{hardcastle07} and \citet{mingo17}. No significant
X-ray diffuse emission is observed in the hotspot complex.

\begin{figure*}
\begin{center}
\includegraphics{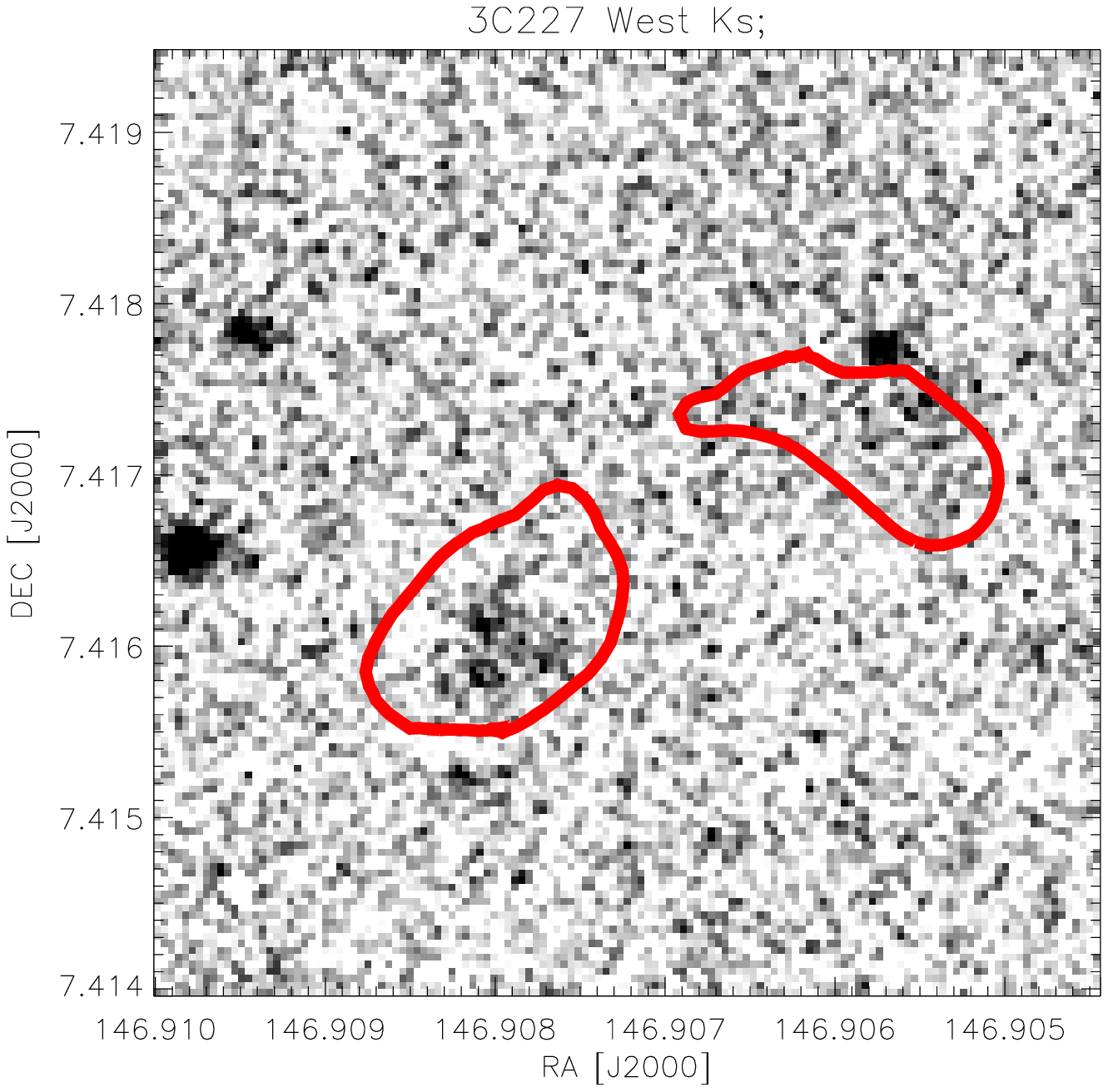}
\includegraphics{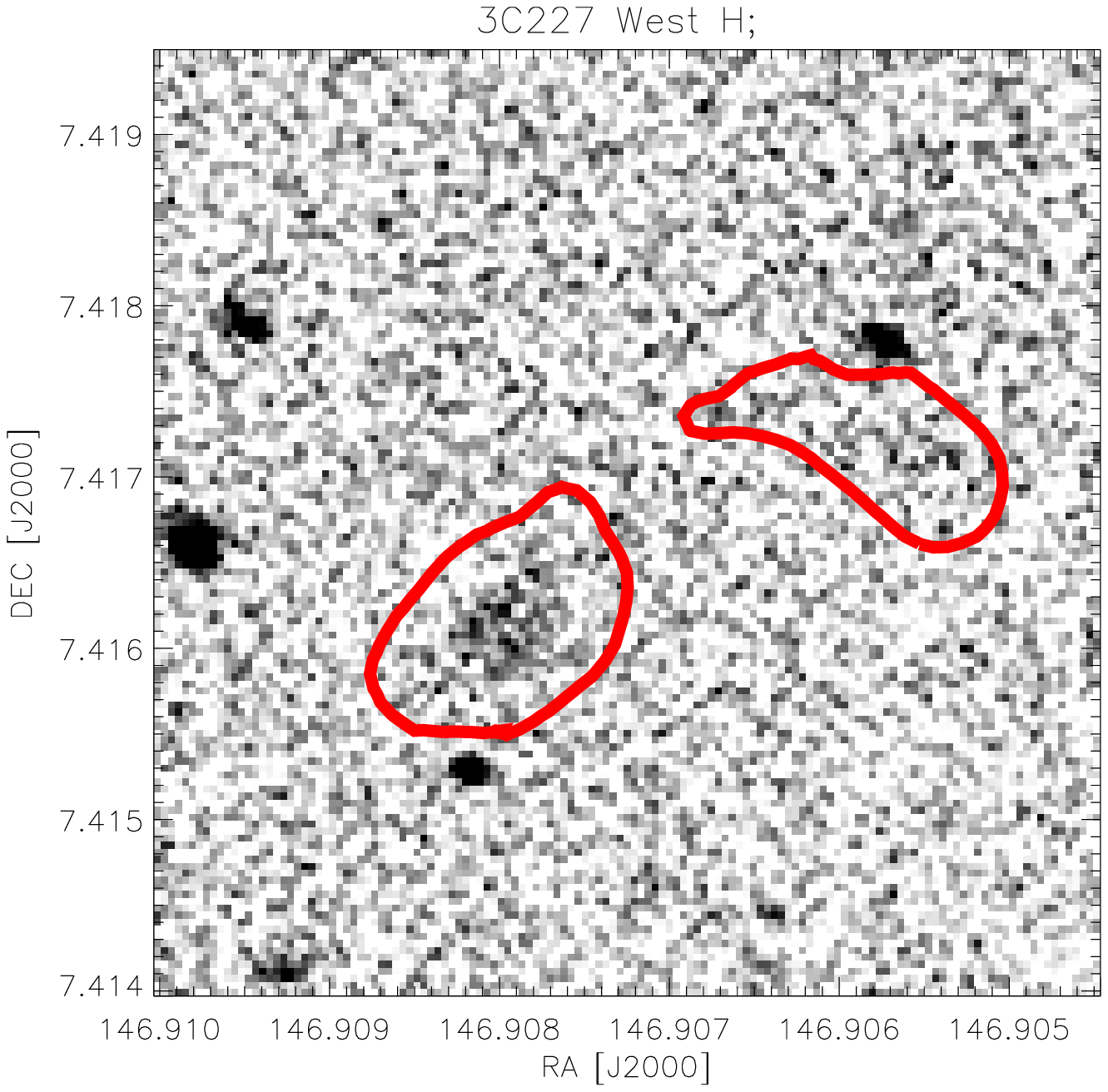}
\includegraphics{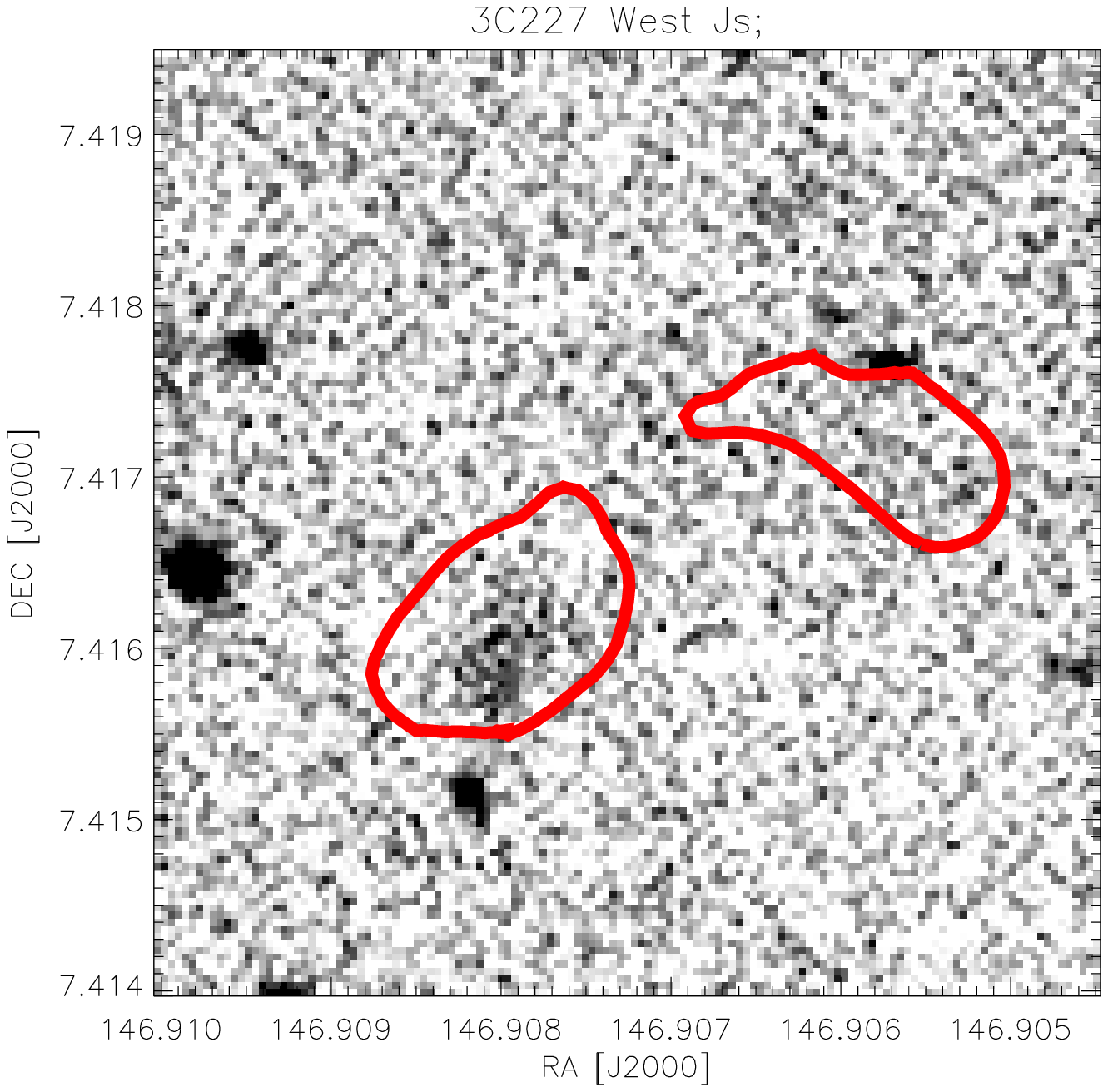}
\includegraphics{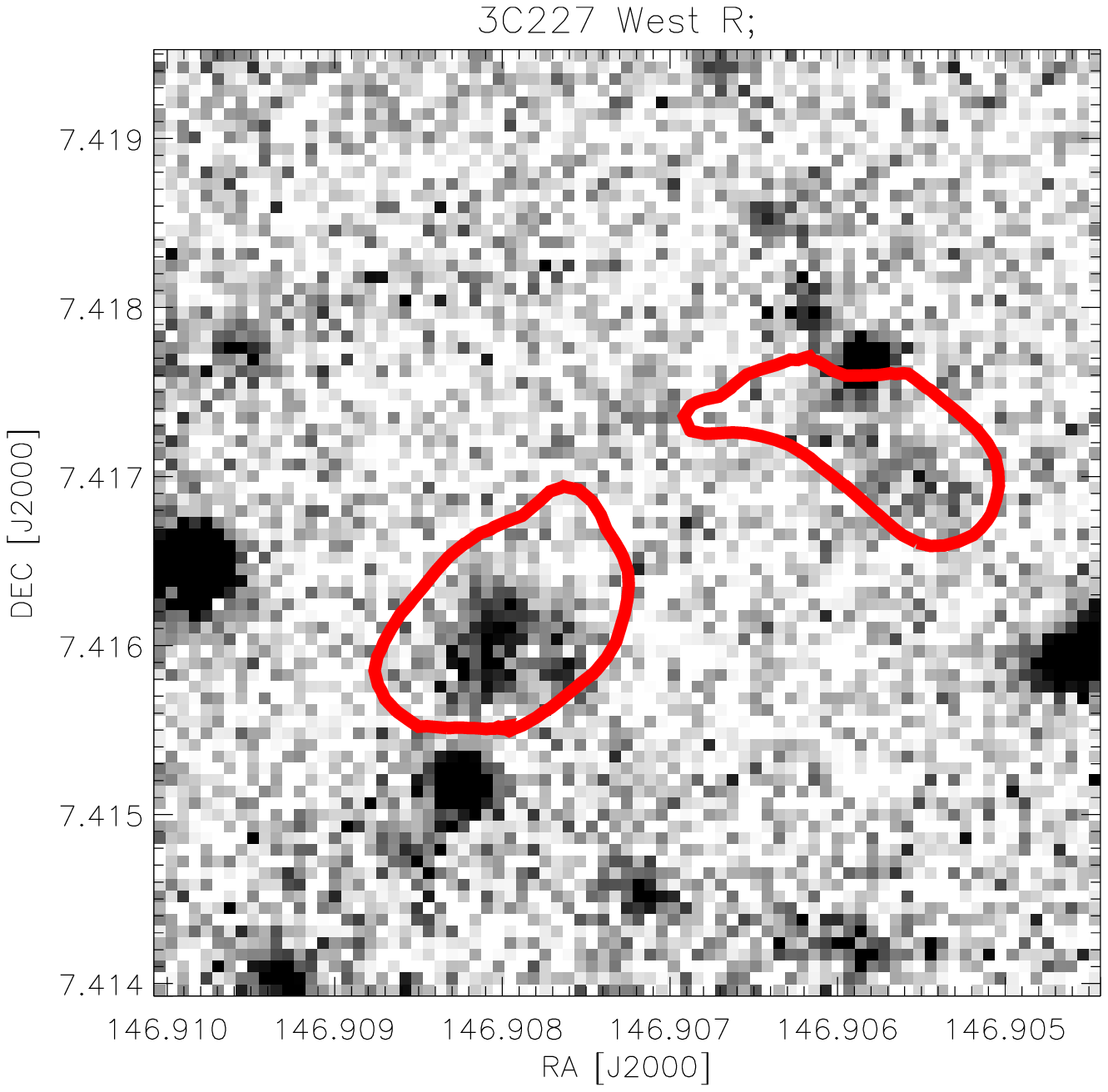}
\includegraphics{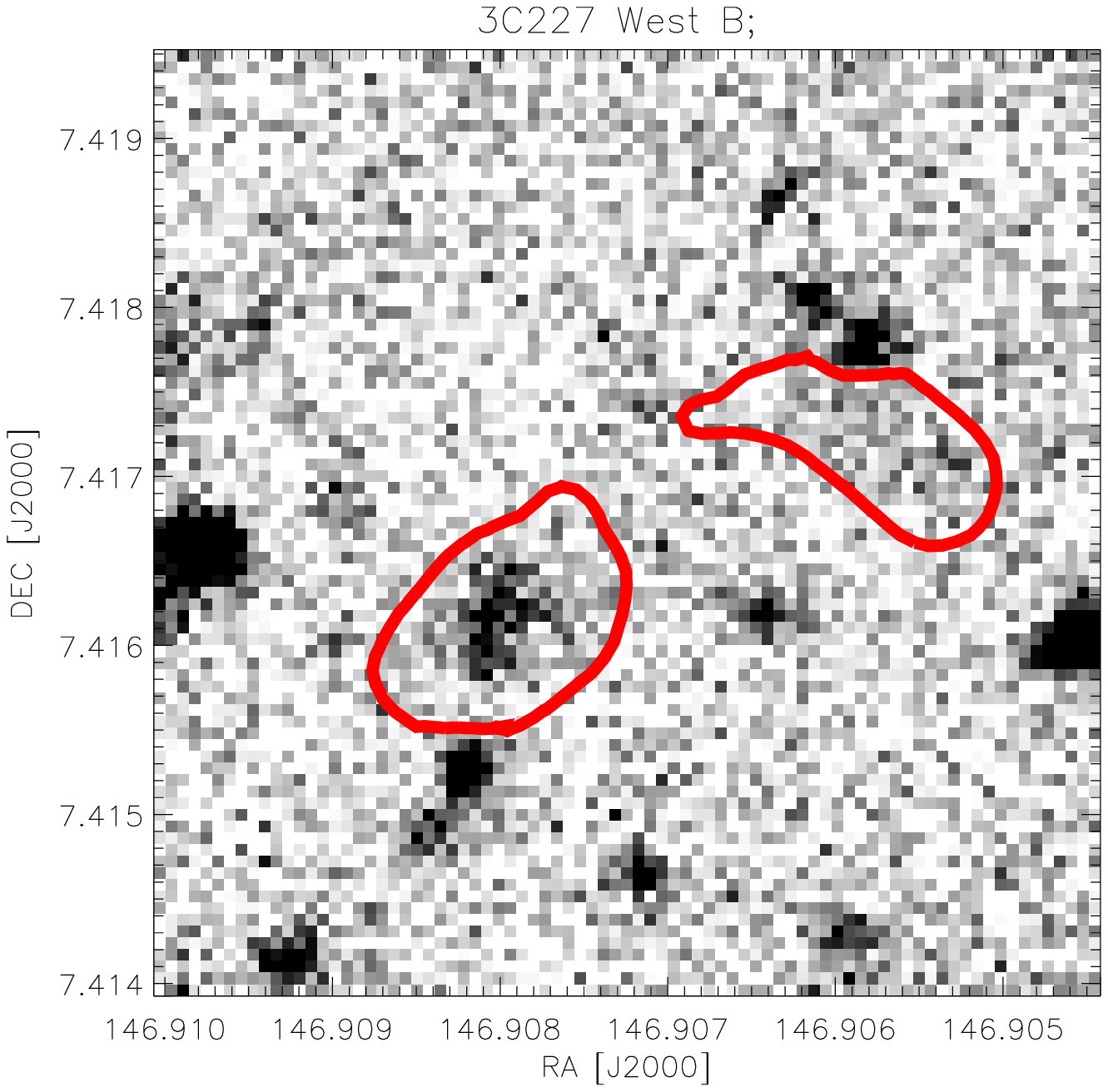}
\includegraphics{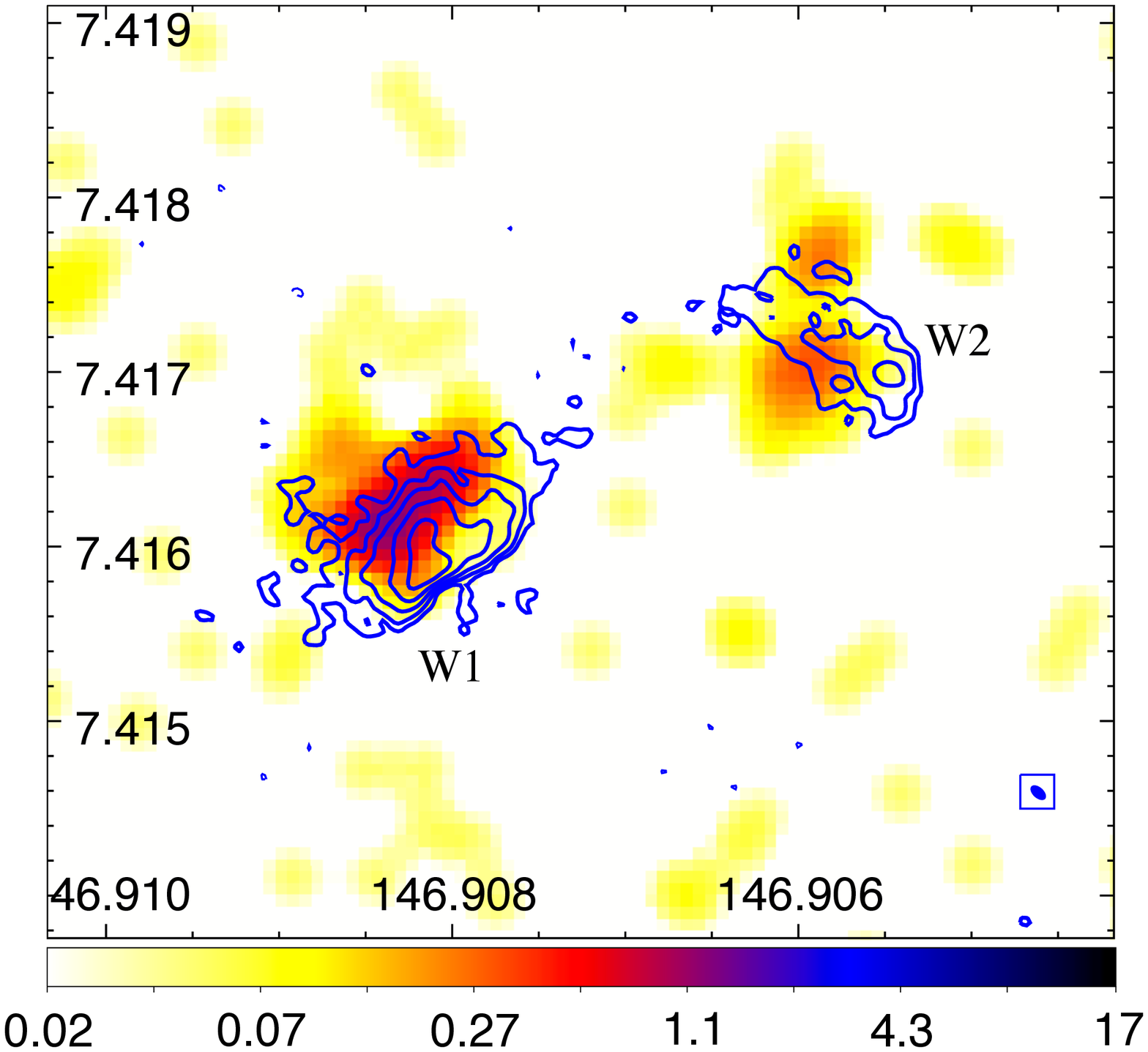}
\vspace{12cm}
\caption{Multi-band images of 3C\,227 West. From left to right
  and top to bottom: VLT Ks, H, Js, R, and B band, and {\it Chandra} X-ray
  images. In the NIR/optical images the contours are the extraction regions
considered for flux measurement and correspond to the 5 per cent of
radio peak flux of each component with the exception of the W2 in
which the region in some bands have been slightly modified from those
shown in the 
figure to avoid contamination from a foreground star. The
intensity scale of the 0.5$-$7 keV image (in count/pixel units) is logarithmic and the image was binned to half the
native pixel size (0.246 arcsec) and smoothed with a
Gaussian using $\sigma$ = 4. 
The radio contours at 8.4 GHz are overlaid on the X-ray image.
The first radio contour represents three
times the off-source noise measured on the image,  while dashed contours indicate the first negative contour ($-$3$\times$rms).  Radio contours increase by a factor of 2.  The radio beam is shown in the box in the bottom right corner of the X-ray image. }
\label{3c227west_figure}
\end{center}
\end{figure*}

\subsubsection{3C\,227 West}
\begin{figure}
\begin{center}
\includegraphics{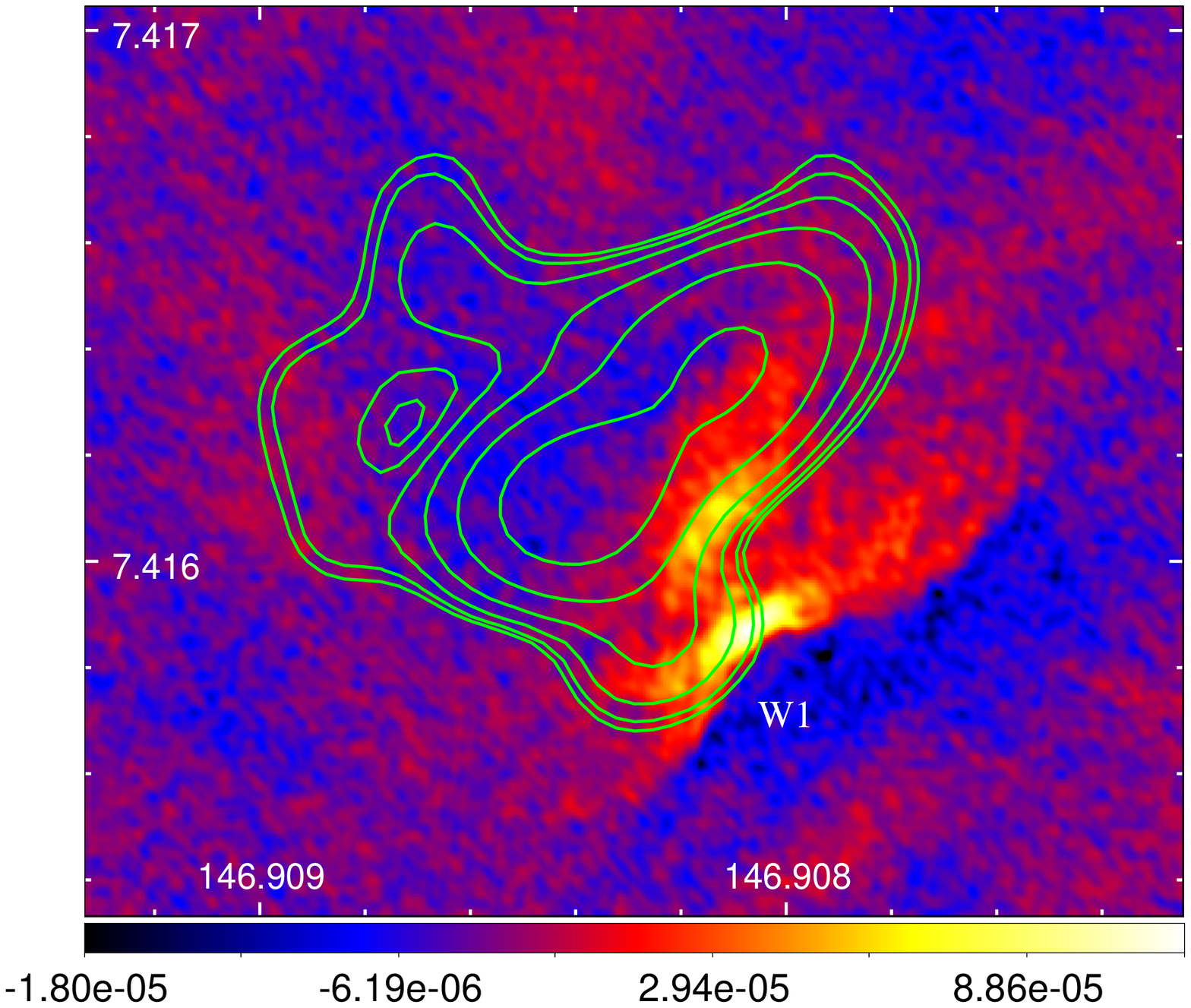}
\vspace{7cm}
\caption{ The JVLA image at 22 GHz of 3C\,227 W1 obtained with natural weighting. A square-root colour scale in units of Jy beam$^{-1}$ is used. The off-source noise level is 6 $\mu$Jy beam$^{-1}$  (see Orienti et al. 2020 for details). The overlaid X-ray contours are in logarithmic scale and start at 6 times the rms (0.005 cts/pixel). The original X-ray image was binned to half the native pixel size (0.246 arcsec) and smoothed with a Gaussian using $\sigma=$4.}
\label{3c227_22GHz_Xray}
\end{center}
\end{figure}

The hotspot complex of 3C\,227 West shows a primary eastern component (W1) and secondary one (W2), located $\sim$10 arcsec west of W1. 
NIR K-band emission from both hotspots
of 3C\,227 West was reported by 
\citet{mack09}. Our multi-band VLT observations detect for the first time the hotspots in
NIR H- and Js-band, and confirm the previous observation in optical R- and B-bands
(Fig. \ref{3c227west_figure}). 
 Because of a foreground star in the proximity of the secondary hotspot, it was necessary to slightly modify the extraction region of the optical flux. For this reason, the measured NIR-optical flux should be considered as a lower limit.
In both hotspots, the NIR/optical
emission is extended and the size is about 1.5$\times$2 arcsec$^{2}$
(3.2$\times$2.4 kpc$^{2}$) for W1 and 
about 1$\times$2 arcsec$^{2}$ (1.6$\times$3.2 kpc$^{2}$) for W2. In particular, in the H-band image the structure of W1 appears resolved both in the S-E to N-W and N-E to S-W directions.
Differently from
3C\,105 South, we did not observe any diffuse NIR/optical bridge connecting the
primary and secondary hotspots.\\
{\it Chandra} observations detected
significant X-ray emission from both the primary and secondary hotspots. As first reported in \citet{hardcastle07}, a 0.8$\pm$0.1 arcsec
displacement to the North (1.3$\pm$0.2 kpc) is
observed between the X-ray emission 
and the radio-to-optical emission of the primary hotspot, with the former occurring upstream towards the nucleus and likely marking regions of current acceleration (Figure \ref{3c227west_figure}).
Note that, as discussed by \citet{hardcastle07}, the fact that for W1 the emission is spatially resolved disfavors the possibility of a by-chance alignment of the radio and X-ray emission.
In the 22 GHz image, W1 is resolved in two arc-shaped components, with the northern one apparently leaning against the edge of the X-ray emission \citep[see Fig. \ref{3c227_22GHz_Xray} and][]{mo20}. The total 22 GHz flux is 27.4\er0.8 mJy.\\ 
\indent
The association of the X-ray emission with the secondary hotspot W2 is instead more uncertain.  The X-ray emission is co-spatial but offset ($\sim$1.4 arcsec to the south-east) with respect to the peak of the radio emission at 4.8 GHz. The morphology in the X-ray and 8.4 GHz maps does not match: the radio emission is sandwiched between two X-ray components, one at about
1.7 arcsec (2.7 kpc) to the 
East of the radio peak and the other (7 counts between
0.5--7 keV) to
the North. In particular, this latter one could be associated with the
foreground star and therefore it was not considered in the estimate of the X-ray flux of W2.  \\
The primary hotspot, W1, is the brightest in
all energy bands: the W1 to W2 flux ratio is
S$_{\it W1}$/S$_{\it W2}$ $\sim$3.5 in radio, $\sim$1.2 in NIR, $\sim$2.5 in
optical, and  $\gtrsim$4 in X-rays.

\begin{figure*}
  \begin{center}
    \includegraphics{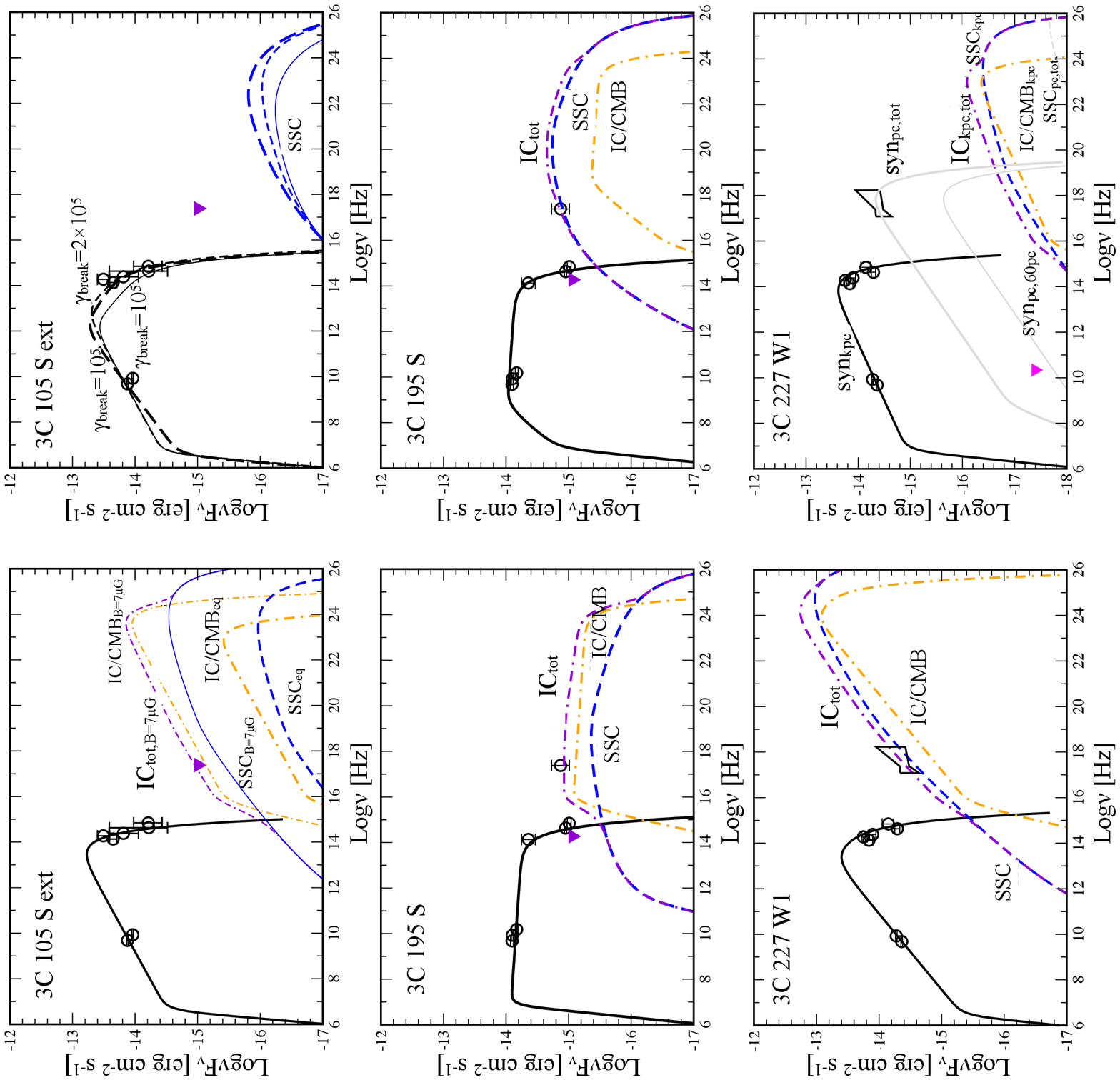}
    \vspace{19.5cm}
    \caption{Observed SED and models of  3C\,105 S Ext, 3C\,195 S and 3C\,227 W1 (see Table \ref{sedtab}). Empty circles are the radio, optical and X-ray fluxes presented in this paper, filled violet triangles are the 3$\sigma$ upper limits. The black solid line is the synchrotron curve, thin and dashed blue lines are SSC curves, dot-dashed orange line is the IC/CMB predicted emission and violet dot-double-dashed line is the sum of the two IC contributions.  Model parameters  have been adjusted to best reproduce the data based on by-eye evaluation.
    Upper left panel: 3C\,105 S Ext data, Model 1 (thick lines) and Model 2 (thin lines). Upper right panel: changes of synchrotron curve for different $\gamma_{break}$. Middle panels: 3C\,195 S observed SED with Model 2 (left panel) and with Model 3 (right panel). 3C\,227 W1: in the lower left panel we show  data with Model 2 and in the lower right panel we show the double-synchrotron component model (see the Discussion). The radio-to-optical emission is modeled with the same parameters of Model 1  (labelled syn$_{\rm kpc}$) and the related total IC emission is given by the double-dotted line (labelled IC$_{\rm kpc,tot}$).  The X-ray emission is produced in compact regions with  $R=60$ pc: the grey thin solid line is the synchrotron emission of a single region  (labelled syn$_{\rm pc,60pc}$). The grey thick solid line is the sum of $\sim$20 compact regions  (labelled syn$_{\rm pc,tot}$). The magenta triangle is the JVLA non detection at 22 GHz at the location of the bulk of the X-ray emission  (Orienti et al. 2020).}
    \label{sed_models}
  \end{center}
  \end{figure*}
    
\section{Broad-band spectral energy distribution}

The new VLT observations, together with the archival data, allow us to model the SED of our targets. 
The NIR-optical data are important to determine the high energy part of the synchrotron spectrum, check for changes in the spectral slope with respect to the radio band and constrain parameters such as  the cut-off frequency (i.e. the synchrotron frequency of the electrons with the largest energy injected in the post-shock region) and the break frequency.
Here, we modeled for the first time the  diffuse emission of 3C\,105 South,  3C\,105 S Ext, whereas SED modeling of its three compact components was presented in \citet{mo12}. The SED of 3C\,195 South in \citet{mack09} did not extend to the X-ray data, which we now included in the modeling.
In  3C\,227 West we focused on the primary hotspot 3C\,227 W1, as the secondary one has less certain multi-band association and suffers of flux contamination in the NIR-optical band. 
 A first analysis of this hotspot in radio and X-rays was discussed in \citet{hardcastle07}. \citet{mack09} provided the first SED  incorporating high angular resolution NIR/optical data. This work improves over it with further high angular resolution IR and optical VLT data. 
 Furthermore, we exploited the spatial and flux information obtained from the
high-angular resolution and high-sensitivity JVLA observations of W1 at 22 GHz, which are presented in a companion paper \citep{mo20}  and summarized here (Sec. 2.3).
We did not model 3C\,227 East as the presence of foreground stars in its hotspot complex precludes an accurate estimate of optical flux from this region. 

We used a leptonic, synchrotron and IC model to reproduce the emission. For simplicity, we assumed a spherical shape of the emitting region.
In case of a different morphology of the observed emission (cylinder, ellipsoid etc.), we calculated the radius $R$ of a sphere with the same volume and uniformly filled by relativistic plasma (i.e. a filling factor equal one was assumed) and magnetic field $B$.
We allowed the value of $R$ to vary within the measured size of the radio emission of each component. 
 The advance speeds measured for the hotspots range between 0.01$c$ and 0.3$c$ \citep[][]{PC03,Nag06,An12}, therefore we began assuming a subrelativistic plasma flow \citep[$v=$0.05$c$ see also][]{kap19}, i.e. values of the bulk Lorentz factor $\Gamma_{bulk}$ close to 1.
The spectrum of the electrons' energy distribution (EED) was described by a single power-law or a broken power-law:
\begin{equation}
N(\gamma)=
\begin{cases} k \gamma^{-p_1}\  \rm for\  \gamma_{min}\leq \gamma<\gamma_{break} \\ 
 k \gamma_{break}^{p_2-p_1}  \gamma^{-p_2}\  \rm for\ \gamma_{break}\leq \gamma<\gamma_{max}\end{cases} 
\end{equation}
where $\gamma_{min}$, $\gamma_{max}$ and $\gamma_{break}$ are the minimum, maximum Lorentz factors and the Lorentz factor at the energy break (or synchrotron frequency break), respectively, and $p_1$, $p_2$ are the EED spectral indexes below and above the break. 
If not differently specified, $\gamma_{min}$ was set to 100. The limited sampling of the SED did not allow us to identify breaks or change of slope in the synchrotron  spectra of our hotspots between the radio and NIR/optical bands, hence a simple power law was typically adopted (i.e. $\gamma_{break}=\gamma_{max}$) and we discuss changes from this initial assumption.
The electrons radiate via synchrotron mechanism; the locally produced synchrotron photons  and photons of the CMB provide the seed photons for the IC mechanism. In the modeling, we began assuming energy equipartition between the particles and the magnetic field.\\
\indent
The values of the model parameters are reported in Table \ref{sedtab}.  Note that our goal here was to achieve a broad evaluation of the models. The values in Table \ref{sedtab} should be considered indicative since, given the limited datasets, we did not perform a fit to the data.\\
\indent
 Only the SED of 3C\,105 S ext (i.e. the diffuse emission), for which we only measured an upper limit in X-rays, could be successfully modeled  assuming an equipartition magnetic field (Model 1). 
For this target, the X-ray upper limit allowed us to estimate the minimum $B$ (assuming no beaming effects), below which the observed radio-to-optical synchrotron emission would imply a detectable level of IC flux in X-rays, $B_{min}\gtrsim$7 $\mu$G (see Figure \ref{sed_models}). 
The magnetic field in equipartition, estimated from the integrated synchrotron spectrum, $B_{eq}=$42 $\mu$G, is compatible with this limit. For $B_{eq}=$42 $\mu$G, the detected NIR/optical diffuse emission is given by particles with $\gamma\approx 10^5-10^6$ and radiative ages of the order of $\approx$17 kyr. Note that, since the emitting region is relatively large ($R\sim$5 kpc), the energy density of the locally produced synchrotron photons is lower than that of the CMB ones, hence the IC/CMB emission dominates over the SSC one.\\
\indent
We explored the possibility that a spectral break is present between the radio and NIR band. In Figure \ref{sed_models} (upper, right panel), we show that for $p_1=$2.6 and $p_2=p_1+1$,  synchrotron curves with $\gamma_{break}$ smaller than $\sim 2\times 10^5$ underestimate the NIR-optical fluxes by a factor $\gtrsim$3.5. This holds true even allowing for  a flatter spectrum, $p_1=$2.5 \citep[corresponding to the radio spectral index of the full hotspot region,  $\alpha=0.75$, in][]{mack09} and $B\sim$35 $\mu$G.\\
\indent
In the other two hotspots,  3C\,195 S and 3C\,227 W1, the NIR-optical emission is resolved but it is compact and not diffuse, as one could expect from a post-shock region. For the assumed bulk motion (0.05$c$), the predicted IC emission under the equipartition assumption (Model 1) significantly underestimates the observed X-ray fluxes.  
If we release the equipartition condition, the X-rays can be ascribed to IC assuming ratios of the energy density of the particles to magnetic field ($U_e/U_B$) larger than $\gtrsim10^3$. In this scenario, the dominant contribution in the X-ray band may be either IC/CMB or SSC (see Figure \ref{sed_models}), depending on the volume of the region. For example, in 3C\,195 S the relative weight of the two contributions is reversed going from the maximum value of $R$ determined by the radio measurements ($R=$2.8 kpc, Model 2 in Table \ref{sedtab}), with IC/CMB$>$SSC, to the assumption of a more compact emitting region, $\sim$1 kpc (Model 3 in Table \ref{sedtab}). This shows that $R$ is a key parameter of the modeling. In X-rays we are limited by the resolving power of the current instruments. However low-frequency observations can now probe the plasma structure down to hundreds/tens of parsecs for the closest targets \citep[see ][ and the JVLA 22 GHz observation of 3C\,227 W1]{mo20}.\\
\indent
An high IC/CMB X-ray flux can be obtained if the plasma in the hotspot is still moving relativistically, with $\Gamma_{bulk}\sim$3--4 ($\approx$0.94$c$--0.97$c$), and is seen at moderate or small inclination angles, $\theta\lesssim$20\dg  (Models 4 and 3 in Table \ref{sedtab} for 3C\,195 S and 3C\,227 W1, respectively). Indeed, nor the scales at which deceleration takes place, neither the bulk flow speeds are known in powerful jets, which could be still mildly relativistic \citep[see e.g.][]{MH09} close to their termination point.
   However, the symmetrical large-scale radio morphology of the two radio galaxies does not support small viewing angles, unless of invoking local deviations of the plasma flow from the direction of the jet's main axis.

\begin{table*}
\begin{center}
\caption{ SED models and input parameters. Columns: 1-model; 2-radius of the emitting region; 3-magnetic field; 4-minimum, maximum Lorentz factors and Lorentz factor at the energy break of the EED; 5-spectral index of the EED below/above the energy break; 6-bulk Lorentz factor; 7-angle between the main jet axis and the observer viewing angle; 8-ratio between the energy densities of the magnetic field and particles.}
\label{sedtab}
\begin{tabular}{lccccccc}
\hline
& $R$  & $B$ &$\gamma_{min}$/$\gamma_{max}$/$\gamma_{break}$ &$p_1$/$p_2$ &$\Gamma_{bulk}$ &$\theta$ &(U$_B$/U$_e$)\\ 
& kpc   &$\mu$G &                                                                                   &                       &                               &deg.      &                       \\
\hline
\multicolumn{8}{c}{{\bf3C\,105 S Ext}}\\
Model 1 &4.9    &42  &100/8e5/--                                                              &2.6/--               &1.0                         &45           &1.0\\     
Model 2 &4.9    &7    &100/2e6/--                                                              &2.6/--               &1.0                         &45           & 0.0013 \\     
\multicolumn{8}{c}{{\bf3C\,195 S}}\\
Model 1  &2.8  &76  &100/1.e6/--                                                        &3.05/--             &1.0                         &45            &1.0 \\                 
Model 2 &2.8  &10   &100/2.0e6/--                                                             &3.05/--             &1.0                         &45           &2.1e-3\\
Model 3 &1.0  &13.5   &100/1.7e6/3.e3                                                   &2.05/3.05        &1.0                         &45           &2.3e-4\\
Model 4  &1.0   &53   &100/7e5/--                                                         &3.05/--             &3.0                         &18.0         &1.0 \\   
\multicolumn{8}{c}{{\bf3C\,227 W1}}\\ 
Model 1   &1.6    &72     &100/9e5/--                                                   &2.6/--                       &1.0                         &45.0       &1.0 \\
Model 2   &1.6    &2.1    &100/5e6/1.5e6                                            &2.4/3.4                 &1.0                            &45.0        &3.1e-6\\                           
Model 3   &1.6    &13     &100/1.5e6/2e5                                            &2.4/3.4                &4.0                             &18.0      &0.15    \\  
\hline
\end{tabular}
\end{center}
\end{table*}

\section{Discussion}
 Low-power hotspots have proved to be optimal targets for studying the synchrotron emission   from the highest energised particles accelerated in the flow: observations in the NIR/optical bands of selected samples have reached  detection rates up to 70 per cent  \citep{prieto02,brunetti03, mack09}. The theoretical explanation is that relativistic particles accelerated in low-power
hotspots, likely with lower magnetic field, have longer radiative
lifetimes and $\nu_{break}$ shifted at higher frequencies (NIR/optical bands) compared with those of high-power hotspots \citep{brunetti03}, which are typically in the millimeter range \citep[e.g.][]{meise97}.

In about 80 per cent of the hotspots in \citet{mack09}
detected in the VLT 
observations (3C\,105 South, 3C\,195 South, 3C\,227 West, 3C\,445
North and 3C\,445 
South) the NIR/optical synchrotron emission either displays compact components surrounded by diffuse emission or an extended structure \citep[see also][]{prieto02,mack09,mo12}.
 
{\it 3C\,105 S ext --} The hotspot 3C\,105 South falls in the first category. In this source, the detection of the optical counterparts of both, the primary and secondary (S2 and S3), hotspots, identifies two main sites of particle acceleration, with the secondary component being likely produced by the
impact of the outflow from the primary upon the cocoon wall \citep{mo12}.
Synchrotron NIR/optical emission enshrouds both components.  Such emission extends at least $\sim$4 kpc (projected size, in H band) to the West of the secondary hotspot (S3), as also seen in the radio structure \citep[see also Figure 1 in][]{mo12}.
If this area coincides with the post-shock region, the detection of optical-NIR emission is somehow surprising in the scenario of one single acceleration episode.
In fact, assuming $B_{eq}$, the electrons responsible for optical-NIR emission have $\gamma>10^5$ and  estimated radiative ages in the range $\approx 10^3$ yr for the compact regions and $10^4$ yr for the diffuse one \citep[see also][]{mack09,mo12}.  Even assuming the optimistic scenario of ballistic streaming of the electrons, for the longest cooling times the particles would cover $\sim$3 kpc, a distance that is barely consistent with the {\it projected} extension of the putative post-shock region. Moreover, a random B field could further increase the path of the electrons in the region, thus making the tension with their radiative lifetimes irreversible.

 One possibility is that the electrons, accelerated in the shock front, stream along the magnetic lines.
If so, then (i) particles are no longer accelerated after leaving the shock region and (ii) the electrons should diffuse (guided by the $B$ topology) on a shorter time than their radiative cooling time, $\tau_{diff}\lesssim \tau_{rad}$  \citep[see e.g. Sec. 4 in][]{meise89}.
 The natural implication of (i) is that  the $\gamma_{break}$ of the electrons producing the diffuse emission ($\gamma_{break,ext}$) cannot be greater than that in the compact regions S2 and S3 ($\gamma_{break,comp}$), and this translates into a constraint on the magnetic field in the diffuse emission region ($B_{ext}$):
\begin{equation}
B_{ext}\gtrsim \nu_{break,ext}\times \frac{B_{comp}}{\nu_{break,comp}}\,\,{\rm G}
\label{B1}
\end{equation}
where $B_{comp}$ is the magnetic field in the compact regions in G,  $\nu_{break,ext}$ and $\nu_{break,comp}$ are the synchrotron break frequencies of the diffuse and compact regions, respectively. A second constraint is obtained from (ii), for dominant synchrotron losses (an assumption that is justified by the results of the SED modeling):
\begin{equation}
B_{ext}\lesssim \left(\frac{2.1\times10^{12} c}{ L_{obs}}\right)^{\frac{2}{3}}\left(\frac{1}{\nu_{break,ext}}\right)^{\frac{1}{3}}\left(\frac{\lambda_{mfp}}{L_{obs}}\right)^{\frac{2}{3}}\,\,{\rm G}
\label{B2}
\end{equation}
where $\lambda_{mfp}$ is the mean-free-path of the electrons at $\gamma_{break,ext}$ (for example the bending scale of $B$), and $L_{obs}$ is the linear size of the diffuse emitting region (i.e. the post-shock region), both in cm. 
Modeling of the SED of 3C\,105 S ext has set a minimum value for $\nu_{break,ext}$ around $\sim10^{12}$ Hz ($\gamma_{break,ext}\sim10^{5}$, see Sec 4 and Figure \ref{sed_models}, upper right panel). The values of $B_{comp}$ and $\nu_{break,comp}$ are taken from modeling of the primary and secondary hotspots, S2 and S3, in \citet{mo12}: $B_{comp}=270-290$ $\mu$G and  $\nu_{break,comp}=(0.75-1.5)\times 10^{13}$ Hz. The size of the region $L_{obs}$ ranges from $\sim$4 to $\sim$10 kpc to account for projection effects. 

\begin{figure}
\begin{center}
\includegraphics{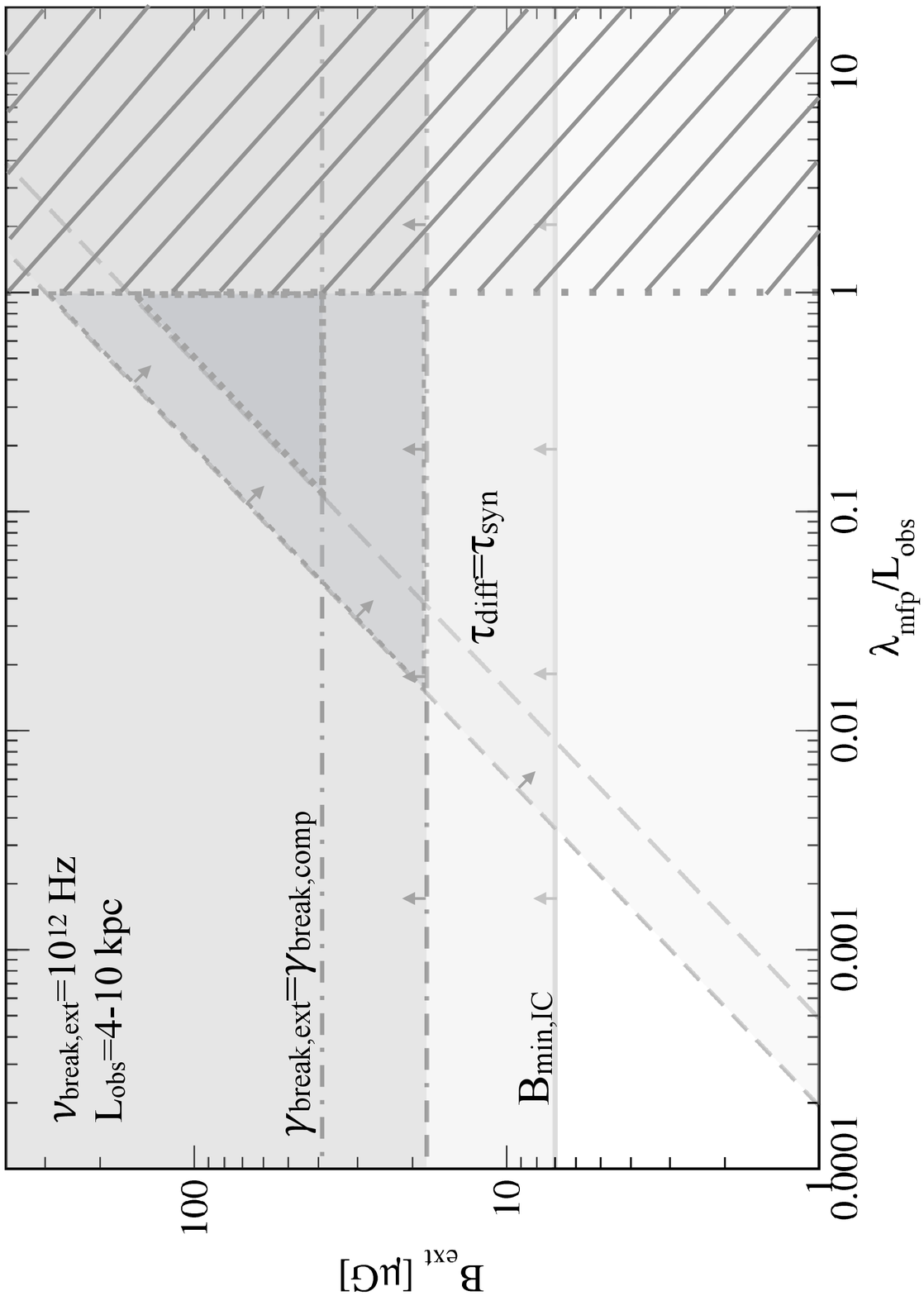}
\vspace{7.2cm}
\caption{ The plot shows the range of allowed values of the magnetic field $B_{ext}$ and $\lambda_{mfp}/L_{obs}$ in the 3C\,105 S ext under the assumption of particles diffusing from the shock-front region along the $B$ field lines without any further acceleration episode. The dot-dashed/dot-double-dashed lines delimit the areas defined by Eq. (\ref{B1}) for the two values of L$_{obs}$. The dashed/long-dashed lines delimit the areas defined by Eq. (\ref{B2}) for the two values of L$_{obs}$.
The solid thick line corresponds to the lower limit given by the non-detection of X-ray emission in 3C\,105 S ext.
The areas of allowed values (for the two values of L$_{obs}$) are given by the two shaded triangles (see Sec. 5 for details).  The $\lambda_{mfp}/L_{obs}>$1 area (striped area) is beyond the putative post-shock region.}
\label{B_lambda}
\end{center}
\end{figure}

In Figure \ref{B_lambda}, $B_{ext}$ is plotted as a function of the $\lambda_{mfp}$ to $L_{obs}$ ratio. 
As a third, less constraining condition, we included in the plot the lower limit on $B_{ext}$, $\gtrsim$7 $\mu$G,  inferred from the non-detection in X-rays (see Sec. 4 and Figure \ref{sed_models}, upper left panel).
The conditions on $B_{ext}$ are fulfilled for $\lambda_{mfp}/L_{obs}\gtrsim$0.01--0.1. For the considered range of $L_{obs}$, $\lambda_{mfp}$  is ($\gtrsim$40--100 pc).
 For a reference, these lower limits of $\lambda_{mfp}$ would be compatible with the sizes of the ordered component of the magnetic field inferred in the hotspots of 3C\,227 and 3C\,445 from the VLA observations at 22 GHz \citep{mo20}. For larger values, instead, the scenario of particles streaming along the magnetic field becomes challenging. High-resolution measurements of the polarized radio component of 3C\,105 S ext could help to probe the $B$ field topology at these scales.\\
\indent
Stochastic (re-)acceleration of the particles out of the main shock site is the other possibility. One can speculate that turbulence, generated via dissipation of the jet's kinetic energy, plays a role, re-energizing particles to maximum $\gamma\sim 10^6$, which produce the NIR/optical emission, but not beyond (hence the non-detection in X-rays).

In the other two hotspots, 3C\,195 S and 3C\,227 W1, the sharp drop of the optical fluxes in the SEDs clearly rules out a single, radio-to-X-ray, component, even allowing for  spectral breaks. The modeling gives hints about the nature of the second radiative component that generates the X-ray emission. 

{\it 3C\,195 S --} The SED of 3C\,195 S is remarkable because of its steep radio-to-NIR spectrum ($\alpha>$1.0). If the spectrum extends below the GHz frequencies without any change of slope, a dominant IC/CMB component at high-energies constrains $\gamma_{min}$ to values $\gtrsim$50--100, not to exceed the observed optical flux. For the same reason, 
if we reduce the volume of the emitting region and the synchrotron radiative field becomes dominant over the CMB one (see Model 3 in Table \ref{sedtab} and Figure \ref{sed_models}), either a $\gamma_{min}$ greater than a few thousands or a spectral break are required.
 If due to radiative cooling, a break at such low energies would imply an extremely old electron population in an hotspot that has switched off, thus excluding fast jet's intermittence. Alternatively, it could reflect an unusual initial EED shape inherent to the acceleration mechanism.
 Interestingly, in a number of hotspots \citep{leahy89,lazio06,godfrey09,mckean16}, evidence for a flattening of the radio spectrum at low frequencies, in the GHz to tens of MHz band, has been found. This could be caused either by a turn-over of the EED \citep[e.g.][]{leahy89} or by the transition between acceleration processes \citep{stawarz07}. 
Observations at low (MHz) radio frequencies sampling the low-energy tail of the synchrotron spectrum \citep[see e.g.][]{hardwood16,hardwood17} can help discriminate among the different scenarios. The field of 3C\,195 has been observed at 150 MHz by the Giant Metrewave Radio Telescope (GMRT) as part of the TIFR GMRT Sky Survey (TGSS) project. We retrieved and inspected the 150 MHz image of our target in the TGSS Alternative Data Release \citep[TGSS ADR\footnote{http://tgssadr.strw.leidenuniv.nl/doku.php};][]{intema17}. Unfortunately, the angular resolution of the survey (25\arcsec$\times$25\arcsec) is not sufficient to reliably de-blend the hotspot emission from the lobe component.
A comparison of the X-ray photon index with the steep radio-optical spectral index is a further test of the (single-zone) IC scenario: for example, similar spectral indexes in the two bands would play in favour of the IC/CMB emission. Indeed, deep X-ray observations are necessary to measure the photon index with a sufficient level of precision.

{\it 3C\,227 W1 --} A single zone radiating model was initially applied also to the SED of 3C\,227 W1. 
As for 3C\,195, a large particle dominance is required if the X-ray emission is of IC origin and not relativistically boosted \citep[$U_B/U_e\lesssim 10^{-5}$, see Table \ref{sedtab} and][]{hardcastle07}. 
For this model, the radio-to-optical spectrum and the best fit value of the X-ray spectral index suggest that the IC total emission peaks above $>$10 keV, at $\approx 10^{24}$ Hz (in Figure \ref{sed_models}).
Therefore, we looked for observations of the source in the hard X-ray to $\gamma$-ray band.
The radio galaxy 3C\,227 was pointed twice by the {\it NuSTAR} mission \citep[the Nuclear Spectroscopic Telescope Array,][]{harrison13}, which is imaging the sky in the hard X-rays (3-80 keV band). 
We retrieved and analyzed the public data. A point source is clearly visible in the {\it NuSTAR} image at the location of the AGN. No significant signal is detected at the hot spot position and, given the instrument PSF, flux contamination from the core results in a relatively shallow upper limit  (6$\times$10$^{-14}$ \ergscm), which hampers a meaningful test of the model. In the MeV-GeV band, the sensitivity limit\footnote{The integral sensitivity is evaluated as the minimum flux above 100 MeV to obtain the 5$\sigma$ detection in 10 years of LAT observation in survey mode, assuming a power law spectrum with index 2. See http://www.slac.stanford.edu/exp/glast/groups/canda/\\lat\_Performance.htm} of the {\it Fermi} Large Area Telescope \citep[LAT][]{atwood2009} is $>2-3\times 10^{-13}$ \ergscm. Thus, the $\gamma$-ray predicted IC flux is at the LAT detection threshold at best. In addition, the LAT PSF (0.8\dg at 1 GeV) would not allow us to disentangle the hotspot $\gamma$-ray flux from the possible AGN  contribution. 

 For 3C\,227 W1 the sensitivity of the current instruments does not make it possible a test of the IC model  in the hard X- and  $\gamma$-rays. However the IC scenario has also other observational issues.
In fact, although the multi-band  emission of 3C\,227 W1 is broadly co-spatial, both hotspots of 3C\,227 West show a displacement of $\sim$1.3$-$2.7 kpc between the X-ray and
radio-to-optical centroids, with the former occurring upstream towards the 
nucleus \citep[see also][]{hardcastle07}. Similar misalignments are observed in other hotspots, like
Pictor A West \citep{hardcastle16} and 3C\,445 South
\citep{perlman10,mo12}. As discussed in these works, such offsets can be hardly reconciled with standard SSC and unbeamed/beamed IC-CMB one-zone models \citep[though see recent developments in simulations modeling the non thermal emission of relativistic flows, e.g.][]{vaidya2018}. 
A displacement between the X-ray and radio emission is instead expected in the model proposed by \citet{gk04} of a decelerating flow, in which freshly accelerated relativistic electrons from the fast upstream region of the flow upscatter to high energies the radio photons produced in the downstream, slower region by electrons that have radiatively cooled.  However, this model involves boosting of the high energy emission and fine-tuned jet parameters (e.g. inclination, bulk motion and location of the radiating regions), while offsets are frequently observed.\\ 
\indent
Alternatively, the X-ray emission may be explained in terms of synchrotron emission from a second population of relativistic electrons, possibly produced in an acceleration event spatially and/or temporally separated from that responsible for the radio-to-optical emission.
In the western hotspot of Pictor A, high resolution radio imaging has unveiled structures with maximum estimated linear scales of $\sim$16 pc, which could be the sites where the X-ray emission is produced \citep{tingay08}.
The discovery of X-ray flux variability on month-to-year timescales \citep{hardcastle16} further supports the hypothesis of the X-ray emission originating in compact (sub-parsec), possibly transient regions.\\
\indent
Evidence of the presence of similar sub-regions, with linear size $\lesssim$100 pc, in  low-luminosity hotspots, including
3C\,227 West, comes from the 22 GHz JVLA observations at high-angular resolution and high-sensitivity that we have recently acquired  \citep[see Orienti et al. 2020 and][for similar results in the NIR/optical band]{prieto02, mack09,mo12}.
 Here we use the information on the physical scales of these regions to investigate  the scenario of a synchrotron origin of the X-ray radiation. 
 The observed radio-to-optical emission is accounted for by synchrotron emission from a kpc-scale region under the assumption of energy equipartition (same parameters as Model 1 in Table \ref{sedtab}).  As discussed in Sec. 4, the total IC flux from this component is significantly lower than the flux measured by {\it Chandra}.
 The X-ray emission results instead by summing together the synchrotron contribution produced by a number of pc-scale regions.   
As an example,  in Figure \ref{sed_models},  we modeled the emission of one such compact region, assuming $R=$60 pc (i.e. within the observed radio upper limits, $B=$70 $\mu$G, $\gamma_{min}=10^3$, $\gamma_{max}=10^8$ and again $U_B\sim U_e$). 
To be in agreement with the JVLA observations, the 22 GHz flux density of each single 60 pc region must lie below the 3$\sigma$ noise level (18 $\mu$Jy) measured at the location of the X-ray emission. For these parameters, a cooling break is expected around $\gamma_{break}\sim 8\times 10^7$, implying that this second synchrotron component rapidly drops at $>$10 keV energies. We assumed $p_1=2.4$ (and $p_2=3.4$ above the break), however a harder synchrotron spectrum ($p1=2.0$) would be equally in agreement with the radio and X-ray observational constraints. For the assumed region's parameters about 20 regions are sufficient to obtain the observed X-ray flux (see Figure \ref{sed_models}). 
Indeed, these numbers should be considered as indicative and could change if we modify the parameters and assumptions (e.g. smaller $R$, departures from the energy equipartition assumption). 
 In this scenario, electrons are efficiently accelerated in clumps. Given the fast cooling time of the high-energy particles, the X-ray emission would trace the most recent acceleration sites, while the bulk of the (slowly evolving) radio emission could result from the accumulation over time of the several acceleration episodes.
 
\section{Summary \& Conclusions}
In this work, we presented new NIR and optical data of low-power hotspots, investigated their structure at different wavelengths and modeled their SED. The main results can be summarized as follows:
\begin{itemize}
    \item we confirm the detection in the NIR/optical bands of all targets, with the exception of one uncertain association  (3C\,227 W2 in 3C\,227 West), with the emission being typically resolved;
    \item the radio and NIR/optical diffuse emission in 3C\,105 South \citep[as already reported by][]{mo12}, which surrounds the bright and compact hotspots, likely coincides with the post-shock region. 
    The constraints on the cooling times and on the minimum mean free path ($\gtrsim$40-100 pc) of the particles producing such emission make a robust case  for some kind of mechanism accelerating particles in the post-shock region, e.g. Fermi II shock re-acceleration as proposed in e.g. \citet{prieto02}. Radio observations probing the configuration of the magnetic field in this region could further test this scenario;
    \item in view of its SED, 3C\,195 South is a good candidate to confirm/disprove the IC hypothesis for the X-ray emission. Our modeling showed that NIR/optical data set precise constraints to the SSC or IC/CMB emission, which are testable with (i) low-frequency radio observations with angular resolution down to a few arcseconds or less, such as those that the Low Frequency Array \citep[LOFAR][]{lofar13} is acquiring  in the northern Hemisphere \citep[see the LOFAR Two Metre Sky Survey, LoTSS,][]{Lotts19}; and (ii) tighter constraints to the X-ray spectral slope; 
    \item  we showed that synchrotron emission produced in  compact regions, whose existence is confirmed by JVLA observations at 22 GHz \citep{mo20}, is a viable explanation for the X-rays in 3C\,227 W1. The large Lorentz factors, 10$^7$--10$^8$, of the electrons emitting in X-rays imply that efficient particle acceleration is ongoing in the clumps. 
    \end{itemize}
The targets of our study are representative of standard low-power hotspots. Hence, it is likely that they share the same properties and mechanisms of their class. To make progress, on one hand we need to increase the sample of sources with high-sensitivity and high-resolution radio observations, which are necessary to map the complex, small-scale structure of the plasma and magnetic field. 
On the other hand, for the first time, we have access to time baselines ($\sim$10-20 years) in X-rays that allow us to test the high-energy variability in the context of the discussed scenarios. 
At the same time, simulations connecting the macro physical scales of relativistic jets with the micro-physics of the particle acceleration and radiative processes \citep[][for a review]{marti19} can provide the theoretical framework to decode the physics of hotspots and jets.

\section*{Acknowledgment}
We thank the anonymous referee for reading the manuscript carefully and making valuable suggestions. Based on VLT programs 72B-0360B, 70B-0713B, 267B-5721.
LC is grateful to INAF-IRA for the hospitality during the course of
this project. 
FD acknowledges financial contribution from the agreement ASI-INAF n. 
2017-14-H.0. This work was partially supported by the Korea's National Research Council of 
Science \& Technology (NST) granted by the International joint research 
project (EU-16-001).
The VLA is operated by the US 
National Radio Astronomy Observatory which is a facility of the National
Science Foundation operated under cooperative agreement by Associated
Universities, Inc. This work has made use of the NASA/IPAC
Extragalactic Database NED which is operated by the JPL, Californian
Institute of Technology, under contract with the National Aeronautics
and Space Administration. This research has made used of SAOImage DS9,
developed by the Smithsonian Astrophysical Observatory (SAO). 
This research has made use of software
provided by the Chandra X-ray Center (CXC) in the application packages
CIAO and ChIPS.

\end{document}